\documentclass[fleqn,usenatbib]{mnras}

\usepackage{threeparttable}
\usepackage{multirow}
\usepackage{hyperref}
\usepackage{lineno}
\usepackage{graphicx}
\usepackage{xcolor}
\usepackage{pdflscape}
\usepackage[utf8]{inputenc} 
\usepackage[T1]{fontenc}    
\usepackage{caption}
\usepackage{subcaption}
\usepackage{url}
\usepackage{amsmath}	
\usepackage{txfonts}
\usepackage{setspace}
\usepackage[normalem]{ulem} 

\def \mysep {\sep} 
\def \mysep { -- } 

\def \bphi{\bar{\phi}}

\def \beq {\begin{equation}}
\def \eeq {\end{equation}}
\def \beqn {\begin{equation*}}
\def \eeqn {\end{equation*}}

\def\aap{Astr. \& Astroph.}

\def\aj{Astron.\ J.}

\def\apj{Astrophys.\ J.}
\def\apjl{Astrophys.\ J.\ Lett.}
\def\apjs{Astrophys.\ J.\ Suppl.}

\def\mnras{Mon.\ Not.\ Roy.\ Astron.\ Soc.}

\def\jcap{JCAP\ }

\title[Galaxy morphologies in S-PLUS]{Deep Learning Assessment of galaxy morphology in S-PLUS Data Release 1} 
\author[C.R. Bom \& A. Cortesi et al.]{
C. R. Bom,$^{1, 2}$\thanks{E-mail: debom@cbpf.br (CRB)}
A. Cortesi$^{3}$, 
G. Lucatelli$^{4}$, 
L. O. Dias$^{1}$,
P. Schubert$^{1}$, 
G.B. ~Oliveira Schwarz$^{5}$,
\newauthor
N. M. Cardoso$^{6}$,
E. V. R. Lima$^{4}$,
C. Mendes de Oliveira$^{4}$,
L. Sodre Jr.$^{4}$, 
A.V. Smith Castelli$^{7,8}$, 
\newauthor
F. Ferrari$^{9}$,
G. Damke$^{10}$,
R. Overzier$^{11,4}$,
A. Kanaan$^{12}$, 
T. Ribeiro$^{13}$,
W. Schoenell$^{14}$\\
$^{1}$ Centro Brasileiro de Pesquisas F\'isicas, Rua Dr. Xavier Sigaud 150, CEP 22290-180, Rio de Janeiro, RJ, Brazil\\
$^{2}$Centro Federal de Educa\c{c}\~ao Tecnol\'ogica Celso Suckow da Fonseca, Rodovia M\'ario Covas, lote J2, quadra J, CEP 23810-000,  Itagua\'i, RJ, Brazil\\
$^{3}$Valongo Observatory, Federal University of Rio de
Janeiro, Ladeira Pedro Antonio 43, Saude Rio de Janeiro,
RJ, 20080-090, Brazil\\
$^{4}$Universidade de S\~ao Paulo, IAG, Rua do Mato 1225, Sao
Paulo, SP, Brazil\\
$^{5}$ Universidade Presbiteriana Mackenzie, R. da Consola\c{c}\~ao, 930 - Consola\c{c}\~ao, S\~ao Paulo, Brazil\\
$^{6}$Escola Polit\'ecnica, Universidade de S\~ao Paulo, Av. Prof. Luciano Gualberto, travessa do politécnico, 380
$^{7}$ Facultad de Ciencias Astr\'onomicas y Geof\'isicas, UNLP, Argentina\\
$^{8}$ Instituto de Astrof\'sica de La Plata, CONICET–UNLP, Argentina\\
$^{9}$ Instituto de Matem\'atica, Estat\'istica e F\'isica, Universidade Federal do Rio Grande (IMEF--FURG),  Av. It\'alia km 8 , Rio Grande, RS, Brazil \\
$^{10}$ Instituto de Investigaci\'on Multidisciplinar en Ciencia y Tecnolog\'ia, Universidad de La Serena, Ra\'ul Bitr\'an 1305, La Serena, Chile \\
$^{11}$ Observat\'orio Nacional, Rua General Jos\'e Cristino, 77, S\~ao Crist\'ov\~ao, 20921-400, Rio de Janeiro, RJ, Brazil \\
$^{12}$ Departamento de F\'isica, Universidade Federal de Santa Catarina, Florian\'opolis, SC, 88040-900, Brazil \\
$^{13}$ Departamento de Astronomia, Instituto de F\'isica, Universidade Federal do Rio Grande do Sul (UFRGS), Av.
Bento Goncalves 9500.\\
$^{14}$NOAO, P.O. Box 26732, Tucson, AZ 85726
}

\begin{document}
\maketitle

\begin{abstract}
The morphological diversity of galaxies is a relevant probe of galaxy evolution and cosmological structure formation, but the classification of galaxies in large sky surveys is becoming a significant challenge. We use data from the Stripe-82 area observed by the Southern Photometric Local Universe Survey (S-PLUS) in twelve optical bands, and present a catalogue of the morphologies of galaxies brighter than $r=17$ mag determined both using a novel multi-band morphometric fitting technique and Convolutional Neural Networks (CNNs) for computer vision. Using the CNNs we find that, compared to our baseline results with 3 bands, the performance increases when using 5 broad and 3 narrow bands, but is poorer when using the full $12$ band S-PLUS image set. However, the best result is still achieved with just 3 optical bands when using pre-trained network weights from an ImageNet data set. These results demonstrate the importance of using prior knowledge about neural network weights based on training in unrelated, extensive data sets, when available. Our catalogue contains 3274 galaxies in Stripe-82 that are not present in Galaxy Zoo 1 (GZ1), and we also provide our classifications for 4686 galaxies that were considered ambiguous in GZ1. Finally, we present a prospect of a novel way to take advantage of $12$ band information for morphological classification using morphometric features, and we release a model that has been pre-trained on several bands that could be adapted for classifications using data from other surveys. The morphological catalogues are publicly available.\\
\end{abstract}

\begin{keywords} 
galaxies: fundamental parameters \mysep 
galaxies: structure \mysep 
techniques: image processing \mysep
methods: miscellaneous \mysep
surveys
\end{keywords}

\section{Introduction}
\label{sec:introduction}

Galaxy morphology is the study of the shapes and structural properties of galaxies. Many different classifications of galaxies have been proposed in an attempt to understand the physics behind the differences in galaxy morphologies (e.g. \citealp{Zwicky1940,Vaucouleurs1959,vandenbergh1998} and references therein). Early galaxy classification schemes already recognized that there are many galaxies (then called nebulae) with spiral arms and others with a smooth, elliptical appearance (e.g. \citealt{Herschel1864}). Still today, with the wealth of  modern galaxy classification schemes and galaxy sub-types known (e.g. \citealt{Borne1999}), the dominant classes for luminous galaxies remain the Spiral (Sp) and Elliptical (ETG) morphologies.

Morphological differences often reflect the presence of different stellar populations \citep{Sanchez2007} and stellar kinematics \citep{Edelen1969,Wang2020}, which is connected to the stellar masses  and environments of the galaxies \citep{Calvi2012,Crossett2014,Sarkar2020,Wu2020}. For example, Sp galaxies have disks, implying rotational support, and star formation tends to occur along the spiral arms. They usually have access to a reservoir of gas to maintain their star-formation activity. In contrast, ETG galaxies are pressure-supported systems and have much smoother surface brightness profiles (though the amount of rotational support increases from E4 onwards believed to be the result of mainly minor mergers (e.g. \citealt{Kormendy2009,Naab2009,Forbes2011,Bernardi2019}). 
As a global rule, massive galaxies evolved faster than less massive ones, but this effect is less pronounced when morphology is taken into account \citep{Camps2020,peng2010mass, bamford2009galaxy}. Morphology is thus a key evolutionary factor that allows us to understand galaxy evolution throughout cosmic time  \citep[e.g. ][]{Shao2015,van2001evolution,kraljic2014links, buitrago2013early}.

Sky surveys in three or more bands have become widespread in the last decades, revolutionizing the fields of large scale structure and galaxy evolution. With the advent of the Sloan Digital Sky Survey (SDSS; \citealt{York2000}), studies that previously typically involved just hundreds to thousands of objects are now based on hundreds of thousands or even millions of objects. With the exploration of these large volumes of data at once also comes the need for massive automatically determined parameters and classifications. The methods used to perform galaxy classifications are quite diverse, and include human classifications by specialists \citep{Nair2010,Ann2015} or citizen scientists \citep{lintott2008,lintott2011,Willett2013,Simmons2017}, and numerical algorithms to determine the main morphological parameters of galaxies   \citep{Spiekermann1992,Storrie-Lombardi1992,Walmsley2020}. Nevertheless, all these methods share the common problem that the image quality decreases as galaxies become fainter or more distant, which may compromise the classifications depending on the resolution and sensitivity of the observations (e.g. \citealt{Povic2015}). Although image resolution and quality are fundamental features to infer morphology, novel techniques and  methods based on Principal Component Analysis (PCA; \citealp{Kelly2004,Wjeisinghe2010}) or 
Machine Learning (ML) have been shown to be  powerful tools to improving galaxy classifications \citep{Calleja2004,Yamauchi2005,Huertas2008,Banerji2010,Dominguez2018,Wu2019,Clarke2020, Barchi2020} and, particularly for some state-of-the-art computer vision methods, to potentially mitigate these issues. In the coming years, we expect rapid progress in the use of these latter techniques, as required by the enormous volumes of data expected from, e.g., the Legacy Survey of Space and Time \citep[LSST; ][]{Tyson2002,Axelrod2006} performed by the Vera C. Rubin Observatory and sky surveys with the Nancy Grace Roman Space Telescope \citep{WFIRST}. 
The volume of data involved makes it unfeasible to perform galaxy morphology using citizen science efforts only. For instance, \citet{walmsley2021galaxy} performed a hybrid approach, trying to combine both citizen science and ML.

ML is an excellent tool for analysing astronomical data given its power of extracting useful information from complex and varied data sets and assist in decision-making processes. Besides its use in the aforementioned classification of the morphologies of galaxies, ML has also been used successfully to detect, e.g., gravitational lenses and interacting galaxies and to classify quasars \citep{Freeman2013,Shamir2013,Bom_proc,Bom2017,Ostrovski2017,Ma2019,challenge01,Knabel2020}. In the last decade, a subfield of ML known as Deep Learning (DL) has emerged as the main technique for computer vision applications, including classification, facial recognition \citep{lu2017simultaneous}, speech detection and characterization \citep{abdel2014convolutional,vecchiotti2018convolutional}, object segmentation, music classification \citep{choi2017convolutional}, and medical prognostics and diagnostics \citep{li2018remaining,hannun2019cardiologist}. DL techniques have thus proven useful to a wide range of multidisciplinary fields, including astronomy.

DL is particularly useful for the development of models that can process complex and minimally reduced (or even raw) data from different sources and extract relevant features that can then be effectively linked to other properties of interest. In particular, Deep Neural Networks (DNNs) are high-performance data-driven models that typically exceed humans in classification tasks \citep{challenge01}. In astronomy, a number of recent works have demonstrated that DNNs can be used to identify morphological features in raw images with minimal intervention from humans \citep{10.1093/mnras/stx1492,Farias2020,hausen2020morpheus,lanusse2018cmu,2019arXiv191104320C,2019MNRAS.484.5330J,2019MNRAS.482..807P,2019MNRAS.484.3879P,challenge01,s2019modular,Barchi2020}.

In this paper, we present the morphological classification of galaxies into Sp and ETG using new data from the Southern Photometric Local Universe Survey (S-PLUS; \citealt{mendes_de_oliveira2019}). 
The First Data Release (DR1) of S-PLUS covers the Stripe-82 region that was also studied by the citizen science project Galaxy Zoo 1 (thereafter GZ1) using the SDSS data.  Our main aim is to test the advantage of using all $12$ broad and narrow bands for performing morphological classifications compared to using the standard method based on a small number of just broad bands. We also test the use of state-of-the-art DL architectures for computer vision and evaluate the importance of training using neural network weights  optimized in other data sets.

This paper is organized as follows. 
Section\,\ref{sec:data} presents the data and the definition of the different samples used in the DL process. Sections\,\ref{sec:deeplearningmodels} and \ref{sec:morphometry} present the DL models and the morphological analysis techniques used in this work, respectively. Section\,\ref{sec:training} gives the details of the training and validation process and Section\,\ref{sec:results} presents the results as well as a consistency check of our methods. We show the agreement between our morphologies and the galaxy spectral types obtained from template fitting codes. We also compare our  morphologies with the output parameters (e.g.,  concentration and entropy  ) from the 2D decomposition code \textsc{Morfometryka} \citep{Ferrari_2015}. Finally, Section\,\ref{sec:discussion} presents a summary and discussion of our work followed by concluding remarks.

\section{Data}
\label{sec:data}

\subsection{S-PLUS}

S-PLUS is an optical survey using 12-bands \citep[the so-called Javalambre magnitude system, described in ][]{JPLUS} that includes 5 SDSS-like bands and 7 narrow-bands centered on important stellar features (the Balmer jump/[OII], Ca H+K, H$\delta$, G-band, Mg b triplet, H$\alpha$, and the Ca triplet). The survey's depth is r$<$20 AB mag for the narrow bands and r$<$21 AB mag for the broad bands. S-PLUS is performed with the T80-South Brazilian robotic telescope located at Cerro Tololo Interamerican Observatory, equipped with a 9.2k$\times$9.2k e2v detector with 10 $\micron$ pixels, resulting in a field-of-view of 2 deg$^2$ with a plate scale of 0.55\arcsec\ pixel$^{-1}$. The first public data release of S-PLUS (DR1) covers $\sim 336{\rm \ deg}^2$ over the Stripe-82.

We have used the full catalogue of the S-PLUS Data Release 1 (DR1) that comprises 3M sources including galaxies, quasars and stars in the Stripe 82 region (more details can be found in \citet{nakazono2021discovery}. The Stripe 82 is an area of sky 2.5  degrees wide by 120 degrees long (-1.25 < Dec < 1.25, 310 < RA < 60) along the celestial equator that has been covered several times by the SDSS survey from 2000 to 2008 \citep{Abazajian_2009}. The S-PLUS survey observed the Stripe 82 area, with a slightly larger spatial coverage (-1.5 < Dec < 1.5, 290 < RA < 60). Specifically, we employed two quantities of that catalogue in order to define the galaxy sample: the magnitude in the Petrosian region in $r$-band (r$_{\rm petro}$) and the probability of an object to be a galaxy (prob$_{\rm gal}$). 
The quantity prob$_{\rm gal}$ is an estimate from a random forest model which was trained in S-PLUS data containing galaxies, quasars and stars. The algorithm uses the $12$ magnitudes of S-PLUS, the Kron radius, full width at half maximum (FWHM), the semi-major and semi-minor axes of the object. The random forest library used was from \texttt{scikit-learn}\footnote{\url{https://scikit-learn.org/}}. 
The interval criteria for these two quantities are:
\begin{align}
    {\rm r}_{\rm petro} < 17 ~{\rm AB~mag}
    \qquad \text{and} \qquad  
    {\rm prob}_{\rm gal} \geq 0.6
    \nonumber
\end{align}

\begin{figure}
	\centering
	\includegraphics[width=0.95\linewidth]{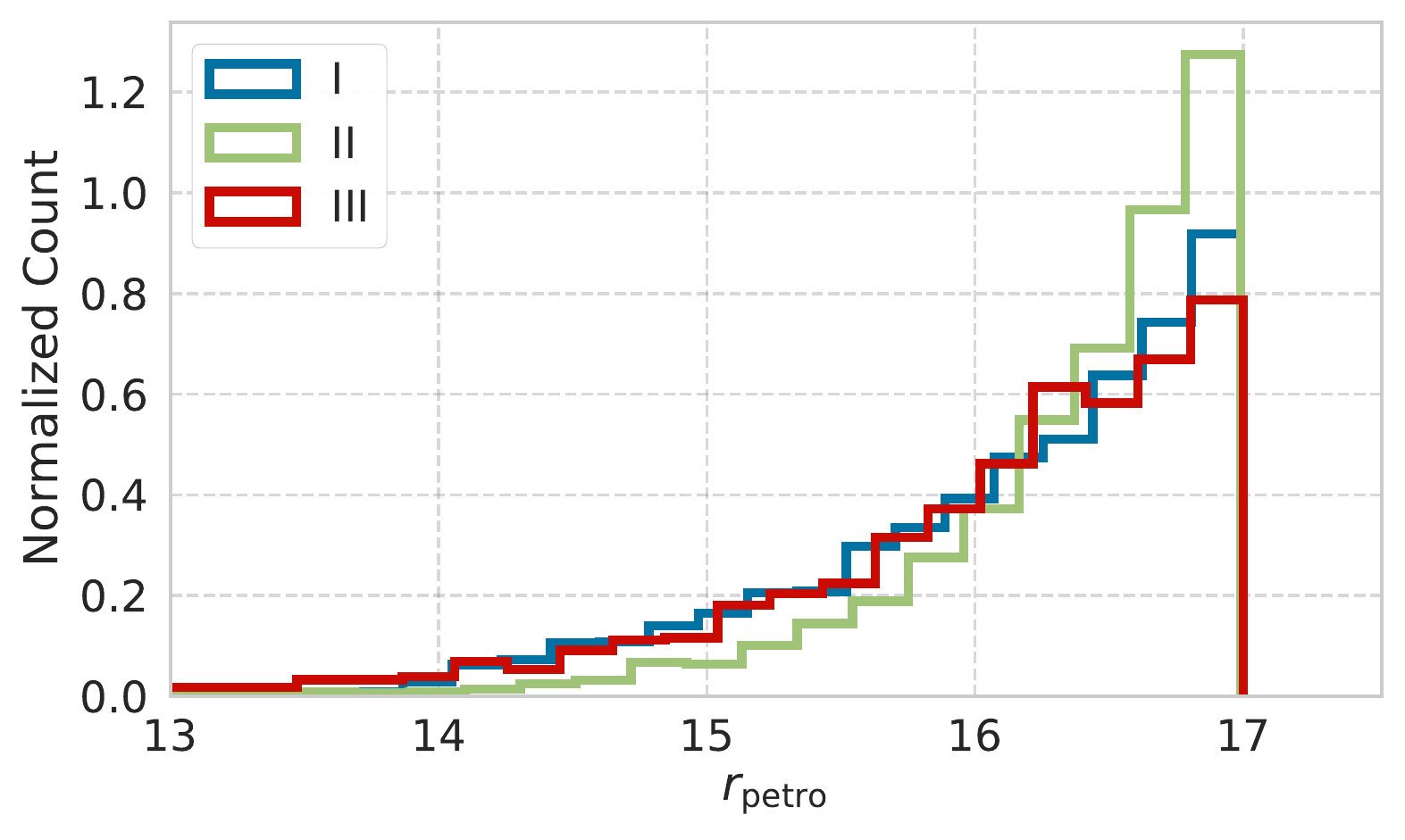}
	\includegraphics[width=0.95\linewidth]{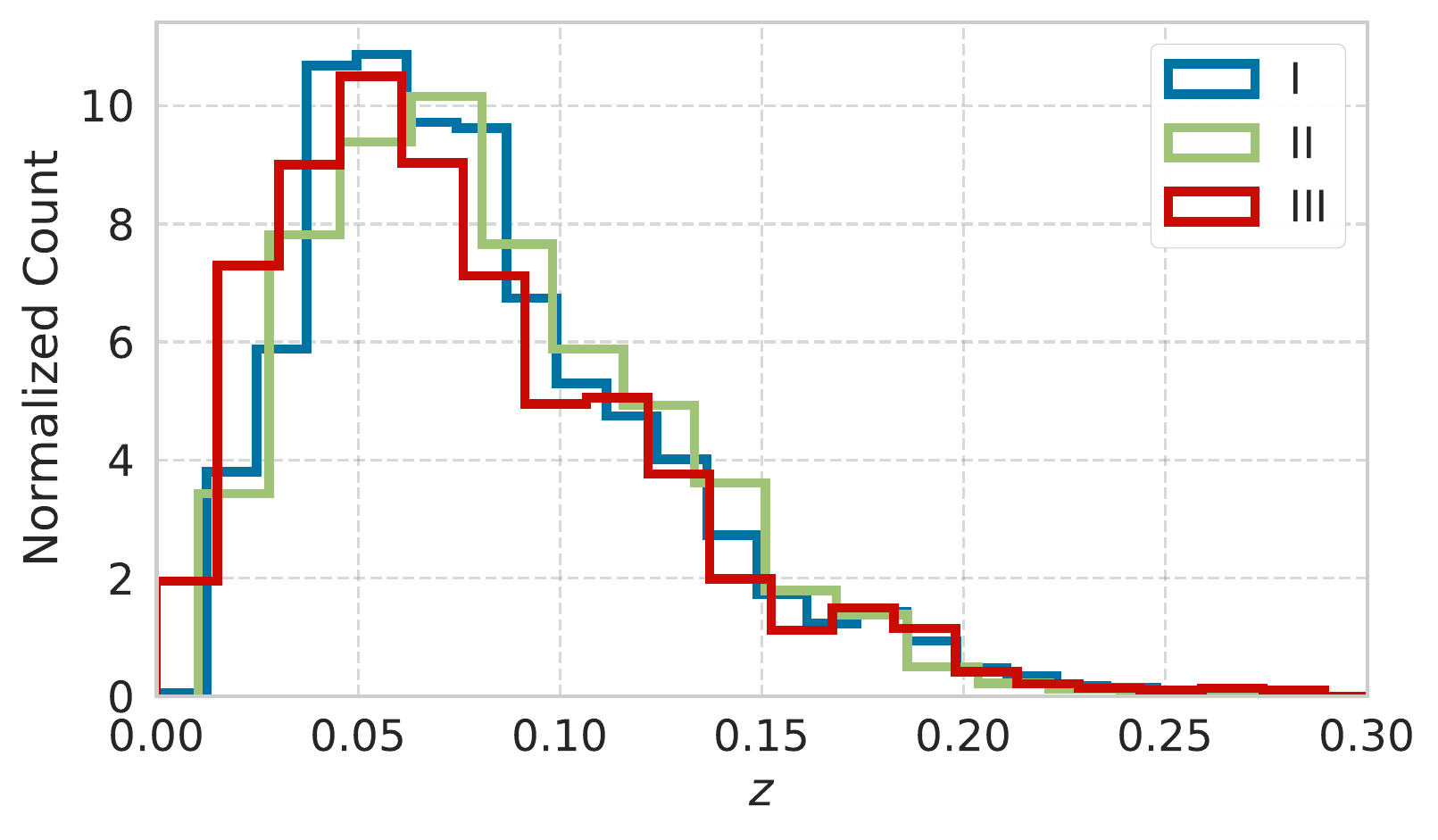}
	\caption{{\it Top:} Normalized histogram of the distribution of the r-petrosian magnitudes (r$_{\rm petro}$) for the three sub-samples considered in our analysis. Sub-sample I is the train and validation set, which are objects classified between ETG and spirals in GZ1. Sub-sample II is the ambiguous sample and galaxies in it are classified as 'uncertain' by GZ1. At last, sub-sample III are objects that are not present in GZ1 and it is our 'blind' set which will be used for test.
    The majority of the galaxies lies in the faint end of the distribution. 
	{\it Bottom:} Distribution of spectroscopic redshift \citep{Molino2020} for the galaxies in the three sub-samples.}
	\label{fig:histsamplesrpetro}
\end{figure}

Following these selection criteria, image stamps for the selected objects were created in the 12 S-PLUS filters for all the 170 fields in DR1. We kept the same standard stamp size for all images of 256$\times$256 pixels$^2$\footnote{The image cutout tasks can be found in this GitHub repository: \href{https://github.com/lucatelli/splus-tools}{https://github.com/lucatelli/splus-tools}.} which corresponds to 140.8 arc-seconds per side. This image size represents a good compromise between the size of the object and the stamp's dimension, to diminish the noise arising from the background in the fitting procedure and the inclusion of neighbouring galaxies. The stamps, in fits format, directly inputted in the pipeline. Thus, we avoid any loss of information due to format conversion or data compression.   

\begin{figure*}
\centering
\includegraphics[width=0.99\linewidth]{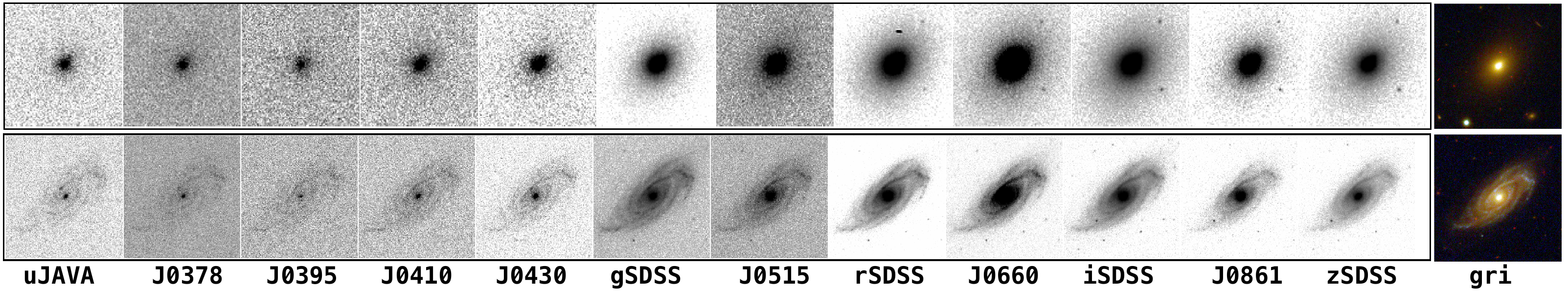}
    \caption{
    \label{fig:example_grid_filters}
    Example of image-stamps showing an ETG galaxy {\it (top panels)} and a Sp galaxy {\it(bottom panels)} in the 12 filters of S-PLUS. The last two panels show the $gri$ colour images.\\
    }
\end{figure*}

As shown in the top panel of Figure\,\ref{fig:histsamplesrpetro}, the majority of the galaxies falls into the faint-end of the brightness regime, assuring that all the visible galaxy light fall within the image size.
Moreover, the DNN algorithm used in this contribution requires the inputs to have the same size. 
Figure\,\ref{fig:example_grid_filters} presents an example of the 12 stamps for an E (top panels) and S (bottom panels) galaxies.

For every image, we also perform a cleaning process in which foreground stars and other objects are removed.  We make use of \textsc{Galclean} \citep{Ferreira_Ferrari_Griffiths_2017,deAlbernaz2018}, which replaces bright sources with the noise distribution of the  background. However, for the DNN method, we still use the original images stamps during training, validation and classification. As in some cases the masked image, resulting from the cleaning process, may contain artifacts (due to the masking procedure) and compromise the performance of the DL analysis. In fact, the DL algorithm may recognise the artefacts as an important element of image-recognition whether the features are quite large compared to the galaxy image, or a similar artefact (as a saturated masked star) is present in several images. Nevertheless, these masks are crucial for the morphometric processing (see Section \ref{sec:morphometry}).

\subsection{Galaxy Zoo 1}
\label{subsec:GZ1_training}
\label{sec:zoo}

GZ1 was a citizen-science project that ran from July 2007 until February 2009 and involved hundreds of thousands of volunteers to visually classify SDSS images in the GZ1 platform \href{https://www.zooniverse.org/projects/zookeeper/galaxy-zoo/}. It allowed not only the classification of nearly 900,000 objects \citep{lintott2008,lintott2011}, but also the discovery of other classes of objects, as, for example, green peas \citep{Cardamone_2009,lintott2008}, and supported a number of follow-up studies, such as those on galaxy spins \citep{Land_2008, Slosar_2009} and mergers detection  \citep{Holincheck2016}.   
In this paper, we use the wealth of data provided for galaxy classification by GZ1, to train a new galaxy Sp-ETG classification for S-PLUS objects. GZ1 is carried out using SDSS DR6 \citep{Adelman-McCarthy2008}, and provide a classification of galaxies into  elliptical, spirals and mergers classes. The median 5-sigma depth for SDSS standard  photometric observations $r = 22.7$ \citep{Adelman-McCarthy2008,Abazajian_2009}.  As from \citet{mendes_de_oliveira2019} the depth of S-PLUS survey in r band is $m_{95\%} = 22.88$.
We note that GZ1 is followed by a successor project, GZ2 \citep{Willett2013}, which classifies galaxies images extracted from the SDSS DR7 Legacy sample \citep{Abazajian_2009}, and from  co-added images of multiple exposures, for the Stripe82 area. GZ2 measures more detailed morphological features than GZ1, such as the presence of bars, spiral arms etc. Moreover, \citet{Jiang2014} release co-added images of the Sloan Digital Sky Survey (SDSS) Stripe 82, reaching a  5$\sigma$ depth  of roughly $23.9$, $25.1$, $24.6$, $24.1$, and $22.8$ AB magnitudes in the five bands, respectively. We leave the exploration of the subtleties features present in GZ2, comparing with deepest data sets as the co-added SDSS data \citet{Jiang2014}, for a future contribution. We selected only galaxies with known spectroscopy (from SDSS DR7), for which the so-called \textit{debiased} classification \citep{bamford2009galaxy} from GZ1 is available \citep{lintott2011}\footnote{Data obtained from \href{https://data.galaxyzoo.org/}{https://data.galaxyzoo.org/}, Table 2, for reference.}. Note that morphological classification is available for all galaxies, with and without spectroscopic redshift determination in Table 3 of  \citet{lintott2011}.

 In fact, a classification bias is present in GZ1 \citep{bamford2009galaxy, lintott2011}, due (i)  to the sample selection (magnitude limits and the small volume at low redshift) and (ii) to signal-to-noise ratio and resolution effects. GZ1 provides four main classes: `Sp', `ETG', `don't know' and `merger'. The classification scheme assumes that galaxies that present spiral arms or are edge-on are classified as Sp while the remaining are considered ETG galaxies. Faint and small objects, though, tend to be classified as early-type galaxies, since spiral features are hard to be identified in such cases. 
 
 \cite{bamford2009galaxy} developed a method to debias the classification for galaxies of known luminosities, sizes and distances, by assuming that their morphological fraction does not vary with the survey depth in bins of galaxy parameter space (i.e. bins of redshift, luminosity and size). To avoid problems due to the low volume at low redshift, the first redshift bin, for which the debiasing technique is applied, starts at z $\geq$ 0.03.
 The morphological classification used as a reference in this paper is the one  published in Table 2 of the GZ1 data release, which provides a debiased likelihood of an object being a spiral or an elliptical galaxy, with a morphology flag, using a binary scheme (0=false,1=true). Details on such classification can be found in Section 4 of \citet{lintott2011}.
 The DL optimization loss procedure used  in this work expects mutually exclusive classes and, it is better constrained to a $1/0$ entry \citep{chollet2018deep}. Early tests presented a high numerical instability when probabilities were used. Thus, we chose for this paper to use the morphology flag instead of the debiased likelihood from Galaxy Zoo. It is important to note that the classification bias in Galaxy Zoo is not due to volunteers' involvement in the study, but to the limitation of the data.  

The merger class, which comprises less than $1\%$ of the classified galaxies and requires a detailed study to consider projection effects \citep{Darg2010a,Darg2010b}, will be studied in a separate paper  (Mendes de Oliveira et al. in prep.). A class that was not considered in the GZ1 project is that of the lenticular (S0) galaxies which, in practice, fall in the S or E classification. Even for an expert, it is hard to disentangle a face-on S0 galaxy from an ETG galaxy or an edge-on S0 galaxy from and edge-on Sp. \citet{bamford2009galaxy} found that the majority of the S0 galaxies fall into the ETG galaxy class, and they suggest that this class is referred to as early-type galaxies. In this work, we will follow this suggestion, leaving to a subsequent S-PLUS paper the challenge of finding bonafide S0 galaxies (Lucatelli et al. in prep.).

\subsection{Deep Learning Sample Definitions}

In this work, we use the debiased Sp and ETG GZ1 samples matched with S-PLUS DR1 as a {\it training and validation set} (sub-sample I); objects that are present in GZ1 and result unclassified due to  the choice of using the morphology flag\footnote{The morphology flag is obtained by applying an 80 per cent threshold on the probability for a galaxy to be in a particular category. Such choice results in  60 per cent of galaxies being unclassified and 3 per cent misclassified. We refer to Figure 5 in  \citet{lintott2011} for a clear visualization of the effect of using different probability thresholds}   and matches S-PLUS DR1, constitute the {\it ambiguous set} (sub-sample II); and S-PLUS galaxies with no counterparts in the GZ1 sample are our {\it blind set} (sub-sample III). In Figure\,\ref{fig:histsamplesrpetro}, we show the histogram  of the distribution of ${\rm r}_{\rm petro}$ and z$_{\rm spec}$ for each sub-sample. More details about these three sub-samples are given below and are also summarized in Table\,\ref{tab:sample_description}.

\begin{table*}
\small
\caption{\label{tab:sample_description} Sample Description. In our sample the data are separated into three groups as indicated below. }
\begin{tabular}{ccl}
\hline\hline
Name & Number of objects & Description  \\

\hline 

I & 4232 & Sp and ETG galaxies matched with GZ1 in S-PLUS DR1 after magnitude cut \\
II & 4686  & Ambiguous galaxies matched with GZ1 in S-PLUS DR1    \\
III & 3274  & Remaining unclassified galaxies in S-PLUS DR1 \\

\hline \hline
\end{tabular}
\\
\end{table*}

\subsubsection{Training and Validation Set - I}

This sub-sample comprises all objects classified by the GZ1 project as `Sp' or `ETG' in Stripe-82 (in total, 4232 objects). We use these objects to train, validate and test the deep learning network. The distribution in magnitudes of this galaxy sub-set is indicated by the blue line histogram of Figure\,\ref{fig:histsamplesrpetro} (top panel).  

\subsubsection{Ambiguous Set - II}
Nearly $52\%$ of the galaxies (4686 objects) in our S-PLUS DR1 sample are classified as {\it ambiguous}. It is worth noticing that in the distribution of the r band magnitude, as shown by the green line histogram in Figure \ref{fig:histsamplesrpetro}, the galaxies are in average fainter than in the other two sub-sets (I and III). Consequently, it is harder to identify the presence of spiral arms, as pointed out by \citet{bamford2009galaxy} and \citet{lintott2011}.

\subsubsection{Blind Set - III}
The third sub-sample (III) represents the S-PLUS data that does not have a counterpart in the GZ1 catalogue.
For example, galaxies without spectroscopic redshift determinations, galaxies falling outside the SDSS Stripe82 area, galaxies with z<0.03 and objects that did not reach the minimum number of votes needed for being classified in GZ1, would compose this sub-sample.
The final number of objects in this set represents $\sim 26\%$ (3274 objects) of the total galaxy sample. We also refer to this sample as \textit{unclassified}.
Since this group is `uncertain' (i.e. it has $\sim 40\%$ probability of being a star),  we made a visual cleaning on the image stamps. We also checked the quality flag \texttt{QF} from Morfometryka (\citealp{Ferrari_2015}; see Section\,\ref{mfmtk}) in order to remove possible non-galaxy objects. In conclusion, only $14\%$ have \texttt{QF} $\neq 0$, which might be caused by, for example, `non-galaxy' objects, foreground stars close to the galaxy, or errors during the Sérsic fit (see \cite{Ferrari_2015} for more details about \texttt{QF}). Such objects, as well as the objects identified as `non-galaxy' during the visual cleaning, have been excluded from this sub-sample, rendering it low contaminated by outliers. Note that this sub-sample's redshift distribution is shifted toward lower redshifts with respect to the other two sub-samples (see Figure\,\ref{fig:histsamplesrpetro}, lower panel). As explained in Section\,\ref{subsec:GZ1_training}, only galaxies with $z\geq0.03$ have been included in GZ1.

\section{Deep Convolutional Neural Networks}
\label{sec:deeplearningmodels}
\subsection{Convolutional Neural Networks for Image Classification}
Convolutional Neural Network (CNN) algorithms are a class of DNN techniques that makes use of a particular type of layers named convolutional layers. These layers were inspired in how animals visual cortex perceive patterns \citep{hubel1962receptive,lecun2015deep}. They consist of kernels, i.e. filters typically represented by matrices, convolved with the data as they flow through the DL model. The DNNs for pattern recognition in images are commonly based on several stacked convolutional layers among with dimensionality reduction layers (or pooling layers) and regularization layers \citep[e.g.,  dropouts or batch normalizations, see][]{Goodfellow-et-al-2016, chollet2018deep}. The trained kernels act as filters specialized in the identification of a specific pattern in the data. Those layers are later connected to traditional fully connected neural network layers that take into account the particular features from the previous layers and outputs the predicted class based on a softmax function \citep{Goodfellow-et-al-2016}. This function returns a vector with the probabilities of the respective potential classes.

\citet{lecun1998gradient} introduced the first widely recognized successful application of these convolutional layers for image identification of handwritten digits. Since that, CNNs have become the primary technique on this field, and its several based architectures won the main computer vision community competitions \citep[see, e.g.][]{ILSVRC15, krizhevsky2012imagenet}, making use of several different CNN based models such as VGG \citep{simonyan2014very}, ResNet \citep{he2016deep} and Inception models \citep{szegedy2015going,szegedy2016rethinking}.

Many patterns of interest in astrophysics can be visually assessed. This spurred the use of CNNs in the field, as the aforementioned examples of pioneering galaxy morphological classification in Section\,\ref{sec:introduction}. One of their major successful visual pattern recognition applications in astrophysical data was the search of Strong Lensing systems. This task has been widely explored both in simulated and real data \citep[see, e.g.][]{Bom2017, jacobs2017finding, lanusse2018cmu, 2019MNRAS.482..807P, 2019MNRAS.484.3879P, Cai2020} and they also have their own computer vision challenges \citep{challenge01,challenge02}, which were both won by CNN-based architectures. It is worth noticing that the model with the best performance in the latest Strong Lensing Challenge its a variation of a family of CNN models known as EfficientNet \citep{efficientnet} that represents the state-of-the-art in one of the most acknowledgeable datasets in computer vision: the ImageNet \citep{deng2009imagenet}. We make use of this kind of model in this contribution as our base model for addressing the morphological classification task.

\subsection{EfficientNet Models}
\label{sec:models}

EfficientNets are a family CNN based models that were scaled and optimized considering a computational resource constrain and accuracy. In its original paper \citet{efficientnet} proposes to start with a model similar to a mobile CNN \citep{howard2017mobilenets}. A multi-objective neural architecture
search \citep{tan2019mnasnet} is performed into this initial model to optimize both accuracy and Floating-point Operations Per Second (hereafter FLOPs). The resulting model is the base model known as $B0$.  This approach, known as Auto Machine Learning (henceforth AutoML), tests several Neural Network hyperparameters combinations to optimize the Machine Learning model, and it is fully described in the previous references. 
After the $B0$ model is defined, \citet{efficientnet} proposed a set of scaling relations to define the width of the layers, i.e. how many kernel units in a layer, the depth, the convolutional blocks and layers and also the resolution of input images. Therefore, constraining the space of DNN hyperparameters to be explored in the optimization processes. 
The intuition for these quantities is that if deeper networks are more prone to learn complex features, on the other hand, they are also more likely to find the problem known as vanishing gradients, i.e., the derivative of the loss with respect to network weights approaches $0$, and the training process is no longer able to minimise the loss function. Also, wider networks and high-resolution inputs are expected to learn fine-grained features \citep{zagoruyko2016wide,howard2017mobilenets}. In this case, the trade-off is that if one defines an excessively wide network, it becomes harder to learn the complex features.
It is worth mentioning that deeper networks are also expected to saturate in accuracy \citep{he2016deep}, thus going deeper might result in unnecessary computational time and other convergence issues.

These scaling relations can be defined as follows:

\begin{equation} \label{eq:scaling} 
\begin{aligned}
\text{depth: } & d  = \alpha ^ \phi \\
\text{width: } & w = \beta ^ \phi \\
\text{resolution: } & r   =  \gamma ^ \phi.  \\
\end{aligned}
\end{equation}
$\phi$ is an integer named compound coefficient.
The coefficients $\alpha$, $\beta$, and $\gamma$ are optimized by a small grid search in the base model, i.e. with fixed $\phi=1$. The number of FLOPS scales with $d \cdot w^2 \cdot r^2$ or, in terms of Equation\,\ref{eq:scaling}, with ($\alpha \cdot \beta ^2 \cdot \gamma ^2)^ \phi$. Thus, to constrain the FLOPs to $2^\phi$, the coefficients $\alpha$, $\beta$, and $\gamma$ are subject to the following constraints:
\begin{equation} \label{eq:optobj} 
\begin{aligned}
& \alpha \cdot \beta ^2 \cdot \gamma ^ 2 \approx 2 \\
& \alpha \ge 1, \beta \ge 1, \gamma \ge 1, 
\end{aligned}
\end{equation}
Under this procedure, one may define a family of EfficientNets from the base model $B0$ to any $BN$ where $\phi=N$. 
It is worth noticing that scaling up an already optimised CNN is a process orders of magnitude less computationally intensive than a fully neural architecture search. This might become prohibitive for many available computer facilities if one has a fixed computation time budget since the number of parameters is of the order of millions or dozen of millions. 

For the Morphological Assessment of galaxies, we made use of the EfficientNet $B2$ model architecture proposed in the original \citet{efficientnet} paper. This network architecture was applied to identification of Strong Lensing \citep{challenge02} winning the II Strong Lensing Classification Challenge. The $B2$ was compared with other EfficientNet models ($B0-B7$) using our baseline input set of g, r, and i bands for morphological classification. The results were similar in terms of accuracy except for the networks with higher numbers of parameters which are the ones more prone to overfit. The $B2$ model was adapted for our problem: the top layer was modified to output the probability of being a Sp or an ETG galaxy. Another significant change was the use of recent state-of-the-art adaptive learning rate optimiser named Rectified Adam \citep[RADAM][]{liu2019variance}. The authors of RADAM have shown empirically that the rectification term applied in the conventional ADAM~\citep{kingma2014adam} optimiser leads to a faster, more stable optimisation which is less sensitive to the choice of hyperparameters such as learning rates. The EfficientNet models are among the highest-ranked in terms of performance for image classification tasks, given an informative (in terms of variety) and sufficiently large structured image dataset as presented by \citet{efficientnet}, outperforming many of the aforementioned computer vision data challenge winners.

\section{Galaxy Morphometry}
\label{sec:morphometry}
Alongside the CNN approach (Section\,\ref{sec:deeplearningmodels}), in this work we performed an additional analysis to recover the morphometric parameters for the three sub-samples and compare their distribution. Such comparison is a consistency check of the CNN classification, since we are able to evaluate whether the same behaviour of the parameters, recovered using Morfometryka \citep[hereafter \textsc{MFMTK}, see][]{Ferrari_2015} in the training and validation set, is present in the {\it ambiguous} and {\it blind} sample. Moreover, galaxies of different morphologies present different values of morphometric parameters \cite[][]{conselice2014} such as concentration and entropy. Thus, the calculation of the distribution of such parameters for the ETG and Sp galaxies classified by the CNN can link the galaxy morphology from computer vision to its structure and physical properties from morphometry. \textsc{MFMTK} and the main morphometric parameters used for this analysis are described in the following sections, while results of this analysis are presented in  \ref{subsec:characterization}.

\subsection{\textsc{Morfometryka}}\label{mfmtk} \ \\  
The morphometric processing is performed using \textsc{MFMTK} \citep[][]{Ferrari_2015}, and it includes photometric (geometric and Sérsic parameters) and morphometric measurements. 
Typical ones among non-parametric quantities are the concentration $C$, asymmetry $A$, smoothness $S$, Gini $G$, second-order moment $M_{20}$, spirality $\sigma_\psi$ and entropy $H$, which constitutes the CASGM-$\sigma_\psi H$ system. These quantities evaluate global and internal properties of galaxy images, such as how asymmetric they are, how concentrated or homogeneous the light distribution is, the existence or not of spiral arms, among other features \citep[for a review see][]{conselice2014}.

All measurements obtained by \textsc{MFMTK} are computed in an automated way from the image and its PSF as inputs. For the analysis and comparison of morphometric parameters, we select the concentration and entropy measurements, since they are more stable in relation to image resolution and signal-to-noise ratio \citep{Ferrari_2015,deAlbernaz2018}. 
A more general study of all  morphometric and photometric parameters recovered with \textsc{MFMTK} is left to another dedicated work, where a more detailed analysis will be conduced using Machine Learning techniques (Lucatelli et. al, in prep).

\subsection{Concentration and Entropy} \ \\
Concentration ($C$) \citep{Kent_1985,Bershady_2000} is widely used to quantify how the brightness is distributed within a galaxy. $C$ is defined as the ratio between two percentile radii of the galaxy. One example is the $C$ index defined considering the radii that contains $80\%$ and $20\%$ of the total luminosity; in that case, $C \propto R_{80}/R_{20}$. In \textsc{MFMTK}, there are different $C$ definitions:
\begin{align}
    C_1= \log_{10} \left( \frac{R_{80}}{R_{20}}\right),\quad 
    C_2= \log_{10} \left( \frac{R_{90}}{R_{50}}\right).
\end{align}

\noindent Generally, ETG galaxies, compact galaxies and classical bulges display higher C values than Sp, disk-like galaxies or pseudo-bulges.

Complementary to $C$, the entropy ($H$) of a galaxy image \citep{Ferrari_2015} quantifies how the light is distributed in the image. The entropy adopted by \textsc{MFMTK} is based on Shannon entropy, 
\begin{align}
H = -\frac{1}{H_{\rm max}} \sum_k^K p(I_k) \log\left[ p(I_k)\right], \quad H_{\rm max} = \log K, 
\end{align} 
where $I_k$ is the intensity value associated with each $k$ bin of the intensity distribution of an image and $p(I_k)$ is the relative frequency of occurrence of $I_k$ of this distribution.
An image with a homogeneous distribution of intensity values will display higher H values than an unequal distribution \citep{Bishop_2007}. In this sense, ETG galaxies may present a lower $H$ in relation to Sp galaxies.

\section{Training and Validation}
\label{sec:training}
\subsection{Preprocessing}
The images were normalized, this is a useful practice for neural network training convergence \citep{Goodfellow-et-al-2016}. A simple image contrast adjustment was performed saturating the bottom 1\% and the top 1\% of all pixel values. We randomly inspect the images to check for quality or issues. In Section \ref{sec:new_catatlogue} we present some panels containing image examples. In the unlabeled sample, we also had to define a  slightly different saturation level as this sample has more galaxies in the faint end and some close to saturated stars.
For training purposes, we made use of data augmentation techniques. This procedure is known to increase DNN performance \citep[see, e.g.,][]{chollet2018deep}. The augmented sample was not used in the validation samples or any network quality metrics, only to enhance the training. To define the augmented sample, we use $180$ degree rotation, horizontal flip and vertical flip.

\begin{table}
\small
\caption{\label{tab:models_desc} Short description of the models used.}
\begin{tabular}{cl}
\hline\hline
Name & Description  \\

\hline 

A & g, r, i bands. No pre-training \\
B & g, r, i bands, Model pre-trained with ImageNet  \\
C & 5 broad bands, g, r, i, u, z. No pre-training  \\
D & 5 broad bands and 3 narrow bands, F515, F660, F861. No pre-training \\
E & 12 bands: 5 broad bands and 7 narrow bands. No pre-training  \\

\hline \hline
\end{tabular}
\\
\end{table}

\subsection{Imbalanced data treatment}
The train/validation dataset is obtained by cross-matching the \mbox{S-PLUS} catalogue with the debiased morphological catalogue of GZ1, as explained in  Section \ref{sec:zoo}. It contains 71\% of Sp galaxies and 29\% ETGs, reflecting the  early-type to spirals ratio at z $\simeq$ 0.03 of the GZ1 datasets. In fact, such ratio is taken as a baseline estimate in the process of debiasing, which assumes that this fraction does not evolve with redshift for bins of fixed luminosity and size (see Appendix A of \citealt{bamford2009galaxy} for a detailed explanation of the debiasing technique). 

\begin{figure}
\centering

\includegraphics[width=0.45\textwidth]{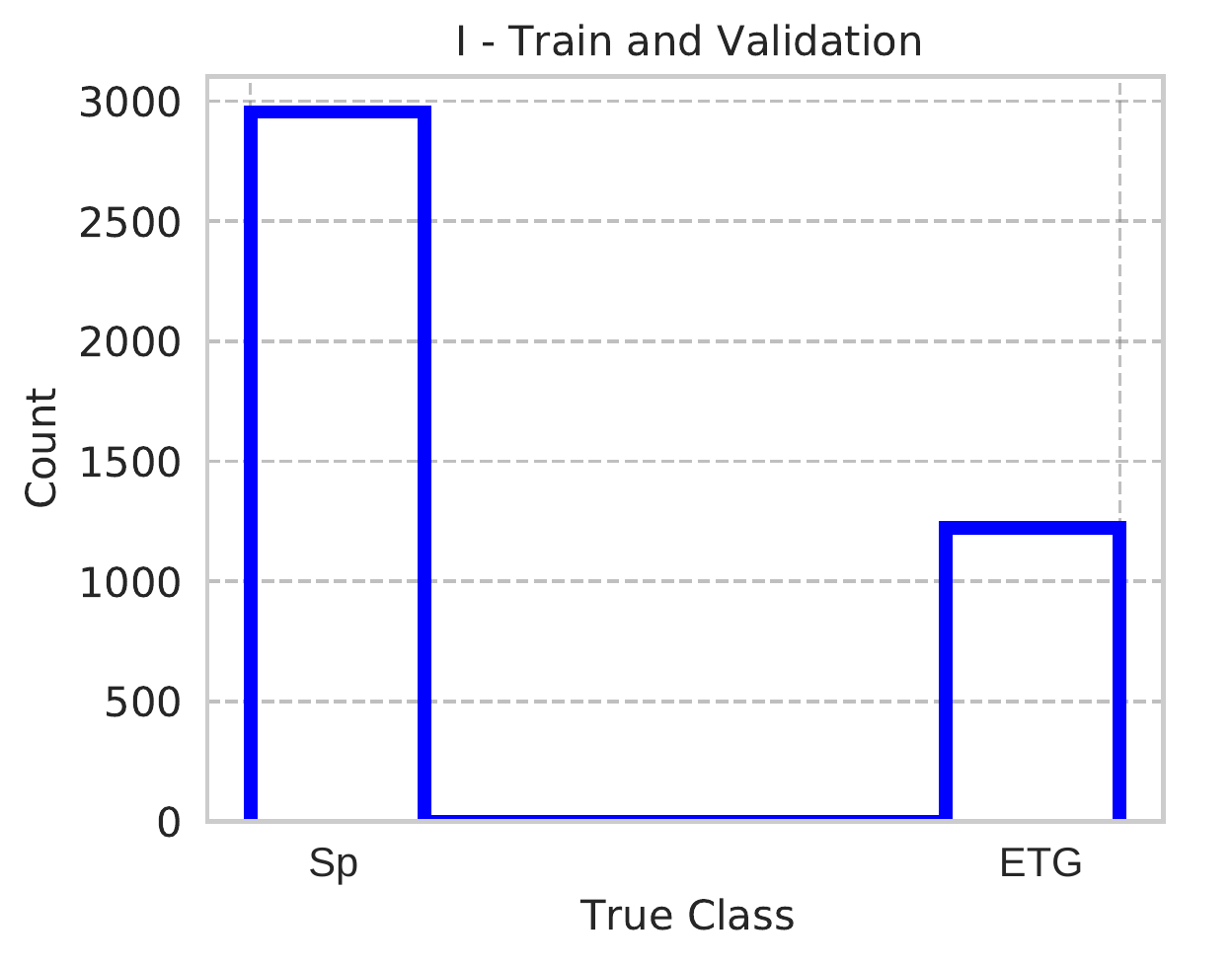}
    \caption{
    \label{fig:HistClass}
    Distribution of galaxy labels in the train/validation sample: i.e. number counts of galaxies classified as ETG or Sp in the unbiased GZ1 data release, present in our galaxy sample from S-PLUS DR1 data release, see Section \ref{sec:zoo}.  
    }
\end{figure}

We present the distribution of both classes in Figure \ref{fig:HistClass}. Due to the data imbalance between the two classes we apply a standard procedure\footnote{see, e.g.,  \url{https://www.tensorflow.org/tutorials/structured_data/imbalanced_data}} to weight each class. We define the weights of a certain class $\alpha$ as:
\begin{equation}
   w_{\alpha}= \frac{N}{m N_{\alpha}},   
\end{equation}
where $N$ is the number of objects in the training set, $N_{\alpha}$ is the number of objects in the class $\alpha$ and $m$ is the number of classes.
The weights are applied in the objective function, so each class has the same impact in the optimization process. Therefore, the procedure prevents the model to bias towards the class with more samples and loose generalization capacity. This is particularly important in problems where one class has orders of magnitude more samples than the other \citep{sun2009classification}. 

\subsection{Training}

We used the dataset I in Table\,\ref{tab:sample_description} described in the previous sections, with the true GZ1 for training and validation purposes. We made use of a standard backpropagation procedure to minimize a cost function, also named as {\it loss}. This function is defined to be a cross-entropy function:
\begin{equation}
H(p,q) = -\sum_{x\in\mathcal{X}} p(x)\, \log q(x),
\end{equation}
where $p(x)$ is the galaxy labelled probability of an object pertaining to a certain class (either 0 or 1 in our case) and q(x) is the probability predicted by our DL method.
We split the dataset in $5$ folds, or subgroups, to perform a cross-validation procedure \citep{moreno2012study}. Therefore we define 5 different splits into a training set, containing 4 folds, and a validation set, containing 1 fold each containing $80\%$ and $20\%$ of the dataset I, respectively. We define the first fold as validation and the other $4$ as the training set for the first set. After the training procedure, we redefined the validation set as the second fold and the others as training. It is worth mentioning that the training subset for each fold is the only one used to update the network's weights in the backpropagation algorithm \citep{ruder2016overview}.   
To decide which is the best network set of weights, we perform 50 epochs and choose the weights defined in the epoch with lower validation loss. The latter would be the weights, to be given to the network, that better generalize the results for a given validation set, unknown to the network training procedure.
The cross-validation method defines one network configuration per training/validation split, which makes possible to evaluate how robust is the training for different sets. Additionally, this method guarantees that each object will be used at least once in the test set.
Using the architecture defined in Section \ref{sec:models}, we set up different models varying the kind of input or the use of preset weights initialization. The models are described in Table \ref{tab:models_desc}.  The model A makes use of three broad bands $g$, $r$ and $i$ in a traditional $3$ channel scheme for image classification in data science problems \citep{deng2009imagenet} using a uninformative weights initialization \citep{glorot2010understanding}. To evaluate the effect of pre-trained weights initialization, we define model B using the same $3$ broad bands. This model also starts with weights derived from pre-training from ImageNet dataset. This idea is known as transfer learning \citep{yosinski2014transferable}. It consists of using a given model (that is, an architecture and its weights) that learned in a given task/data domain, in our case classifying the ImageNet classes, as a base model for another task/data domain. This method relies on the hypothesis that a sufficiently large and well fitted DL model is effective in learning features and some of those features can be shared among tasks like different sets of image classification problems. We let all the weights free to be optimized in our training, which means that we only used the pre-trained weights to initialize the network.
The subsequent models do not use the pre-trained initialization since they are defined in a different number of channels than the original ImageNet. Model C makes use of all $5$ available broad bands $g,~r,~i,~u,\text{ and }z$. Model D is set to use the aforementioned broad bands and also $3$ narrow bands with higher signal-to-noise ratio, namely $F515$, $F660$ and $F861$. These bands can retain information of important stellar features, such as $H\alpha$, Mgb triplet, and Ca triplet. Yet those emission lines  fall into the corresponding filter only till a certain redshift (for example, the  $H\alpha$ line falls into the $F660$ filter only until $z\simeq0.015$). Given the redshift distribution of the galaxies in this sample ($z_{max} \simeq 0.3$, $z_{peak} \simeq 0.05$);  see the lower panel of  Figure\,\ref{fig:histsamplesrpetro}), this is not the case for the majority of the targets, and the differential factor for   the $F515$, $F660$ and $F861$ images is the higher $S/N$ among the narrow band images (see Figure\,\ref{fig:example_grid_filters}). 
Finally, model E makes use of all available broad and narrow bands from S-PLUS (a total of $12$).

\begin{figure*}
\centering
\includegraphics[width=1.0\textwidth]{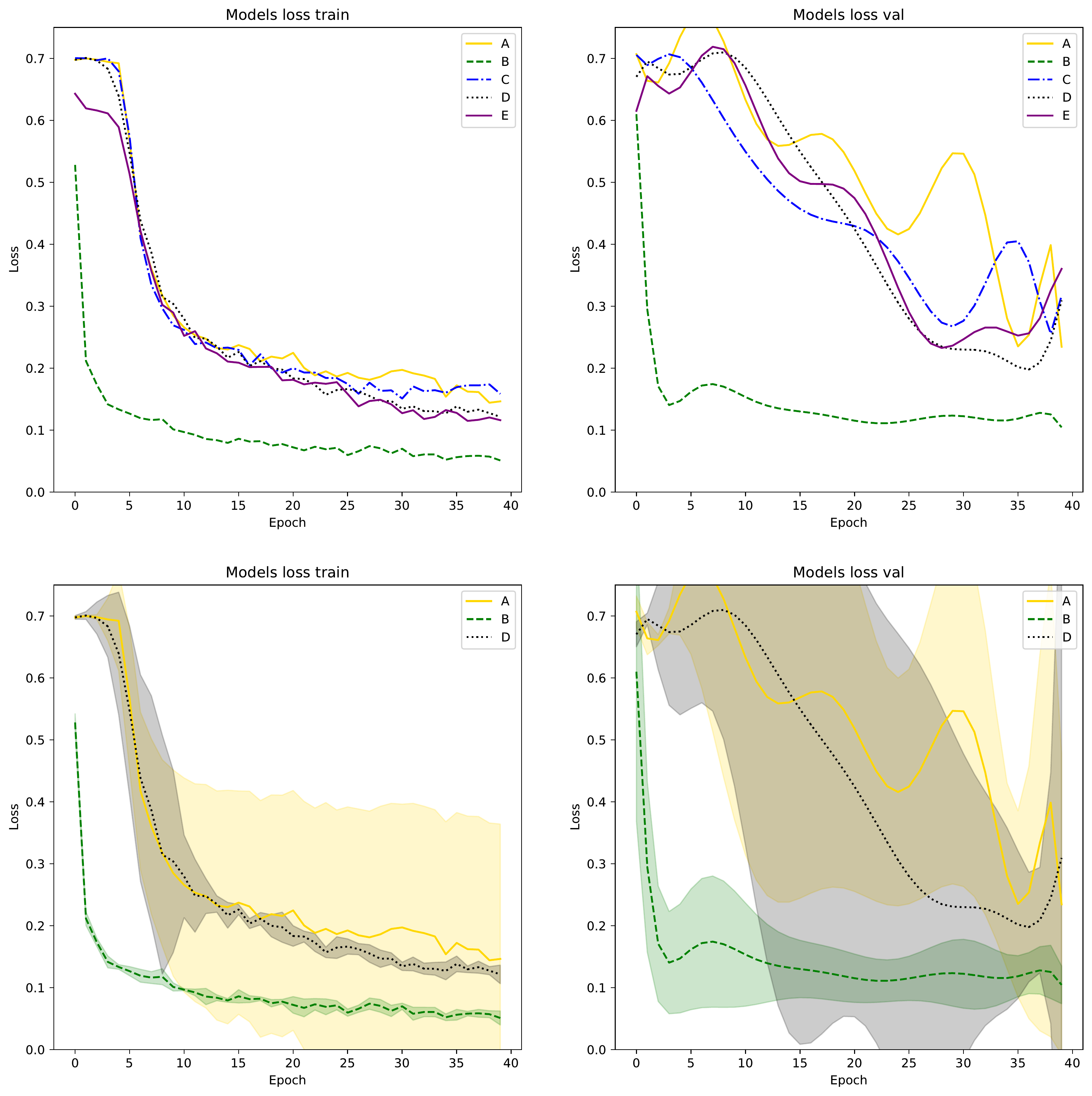}
    \caption{
    \label{fig:loss}
     The objective function results for: A, C, D and E are the results for the 3-,5-,8-, and 12-band models, respectively, all without ImageNet weights. Model B is a 3-band model used ImageNet pre-trained initialization weights. The shaded regions represents the standard deviation in the folds due to cross-validation procedure. 
    }
\end{figure*}

The training plots are presented in Figure   \ref{fig:loss}. The left panel presents the optimisation in the training dataset, and the right panel presents the training performance in the validation dataset. All the models in the training dataset started with loss between $0.6-0.7$ and optimized until reach a phase of stability. However, the model with pre-trained ImageNet weights did perform better in terms of median loss in all folds. The baseline model with $3$ bands presented a high instability in the training process and did not converge to a minimum in all the folds; thus, it presents a high standard deviation for training loss compared to other models. On the other hand, the models with $5$ and $8$ bands had a more stable convergence and performed similarly. It is worth noticing that, regarding the validation loss, none of the models present strong overfitting considering their variability in the folds. However, by looking into individual folds, we observed that the models with $5$ or more bands tend to overfit after $\approx 40$ epochs. For the following analysis, we use the models at the epoch with lower loss in the validation sample.

 It is also worth noticing that all considered configurations, except the pre-trained with ImageNet, optimised the loss function in early epochs. However, the validation loss needed several epochs more to reach a comparable low value indicating that the DNN learned how to generalise for a dataset different from the one used in training (e.g. the validation dataset).

\section{Results}
\label{sec:results}

\subsection{Model Performance}
\label{sec:model performance}
The model performance is evaluated only in the validation sets, i.e. the sets that was not used to optimize the DNN weights during the training.  The uncertainty in each metric is derived by the $k$-folding procedure described in the previous section. We consider as metrics the purity (or precision), completeness (or recall, true positive rate) and false alarm rate (or false positive rate). The precision for the sample of galaxies of a class $\alpha$ is defined as:
\begin{equation}
\text{precision}=\frac{|\{\text{galaxies in $\alpha$}\}\cap\{\text{galaxies classified as $\alpha$}\}|}{|\{\text{galaxies classified as $\alpha$}\}|}.
\end{equation}
The recall can be defined as:
\begin{equation}
\text{recall}=\frac{|\{\text{galaxies in $\alpha$}\}\cap\{\text{galaxies classified as $\alpha$}\}|}{|\{\text{galaxies in $\alpha$}\}|}.
\end{equation}

\noindent The false alarm rate is:
\begin{equation}
\text{false alarm}=\frac{|\{\text{galaxies not in $\alpha$}\}\cap\{\text{galaxies classified as $\alpha$}\}|}{|\{\text{galaxies not in $\alpha$}\}|}.
\end{equation}

\noindent The purity is the percentage of the elements classified as a given class that pertains to that class. The recall is the percentage of elements classified as a given class that pertains to that class compared to all existing class elements or how complete the sample is. The false alarm rate is  how many objects classified as a given class are wrongly classified compared to all elements that do not pertain to that class.
As the DNN outputs a number representing each class's score (ETG or Sp), to obtain a precision and recall it is necessary to define a threshold $t$. If a given probability of a class in our validation dataset is above the threshold ($p>t$) we consider the object as part of this class. To define an optimized threshold and also to assess the performance, we plot the precision and recall and the Receiver Operating Characteristic (ROC, i.e. recall and false alarm rate) curves for any threshold in the range $[0,1]$. The Area Under the Curve of the ROC curve (AUC) is an intuitive measurement of classification performance: a perfect classifier would have ${\rm AUC}=1$. In contrast, a random choice in a balanced dataset would have ${\rm AUC}=0.5$. The perfect classifier is the one that approaches the point $(0,1)$, i.e. $0\%$ false positives for  $100\%$ completeness.

\begin{figure*}
\centering
\includegraphics[width=1.0\textwidth]{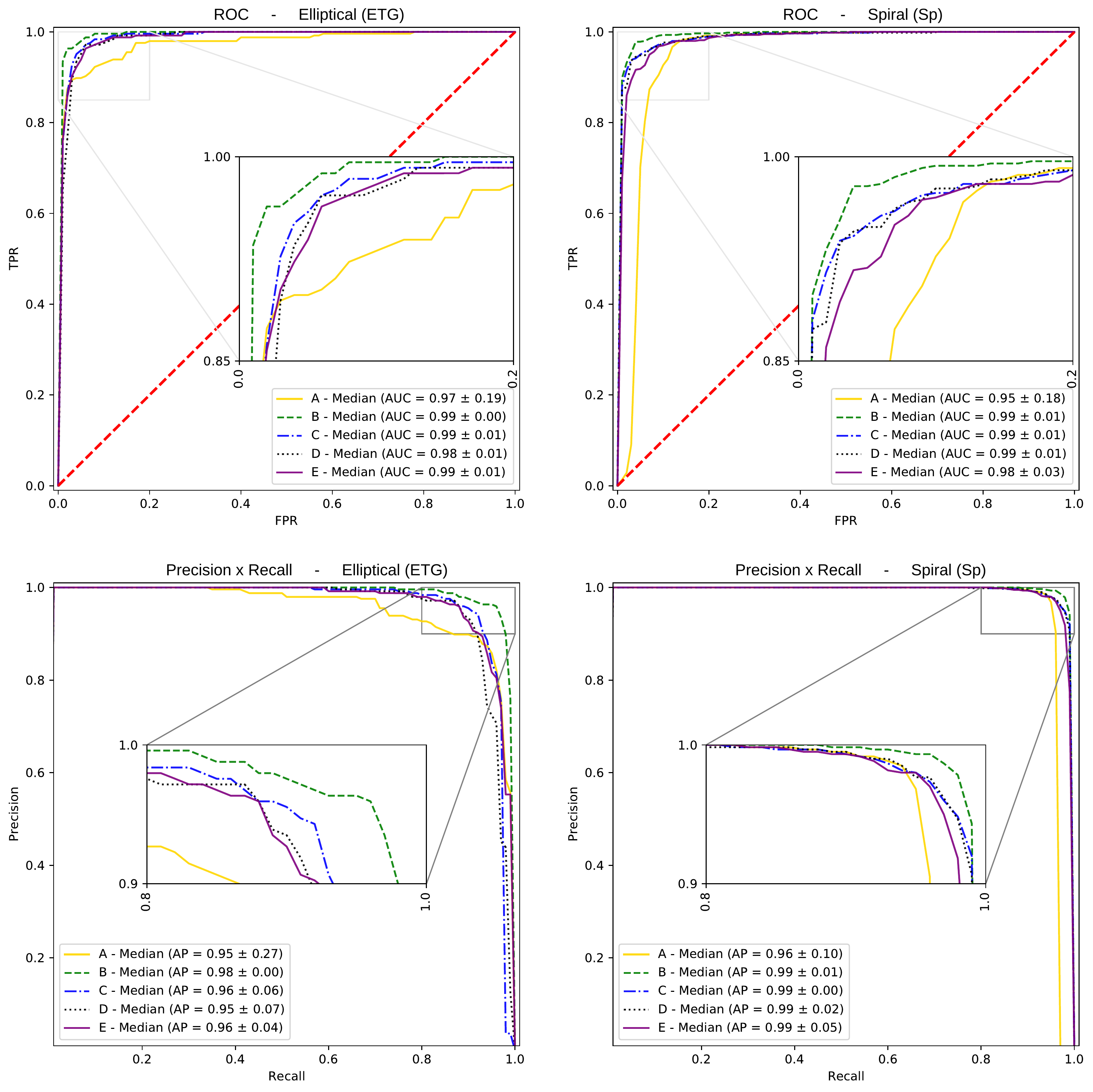}
    \caption{
    \label{fig:MC_EfficientNetB2_1}
    Results for A, C, D and E for the 3-, 5-, 8- and 12-band models, respectively. Model B used pre-trained ImageNet weights initialization with $3$ bands. The upper panel presents the Receiver Operating Characteristic curve (ROC) where TPR is the True Positive rate (completeness or recall) and FPR is the false positive rate. The lower panels presents the Precision (purity) and Recall (completeness).
    }
\end{figure*}

\begin{figure*}
\centering
\includegraphics[width=1.0\textwidth]{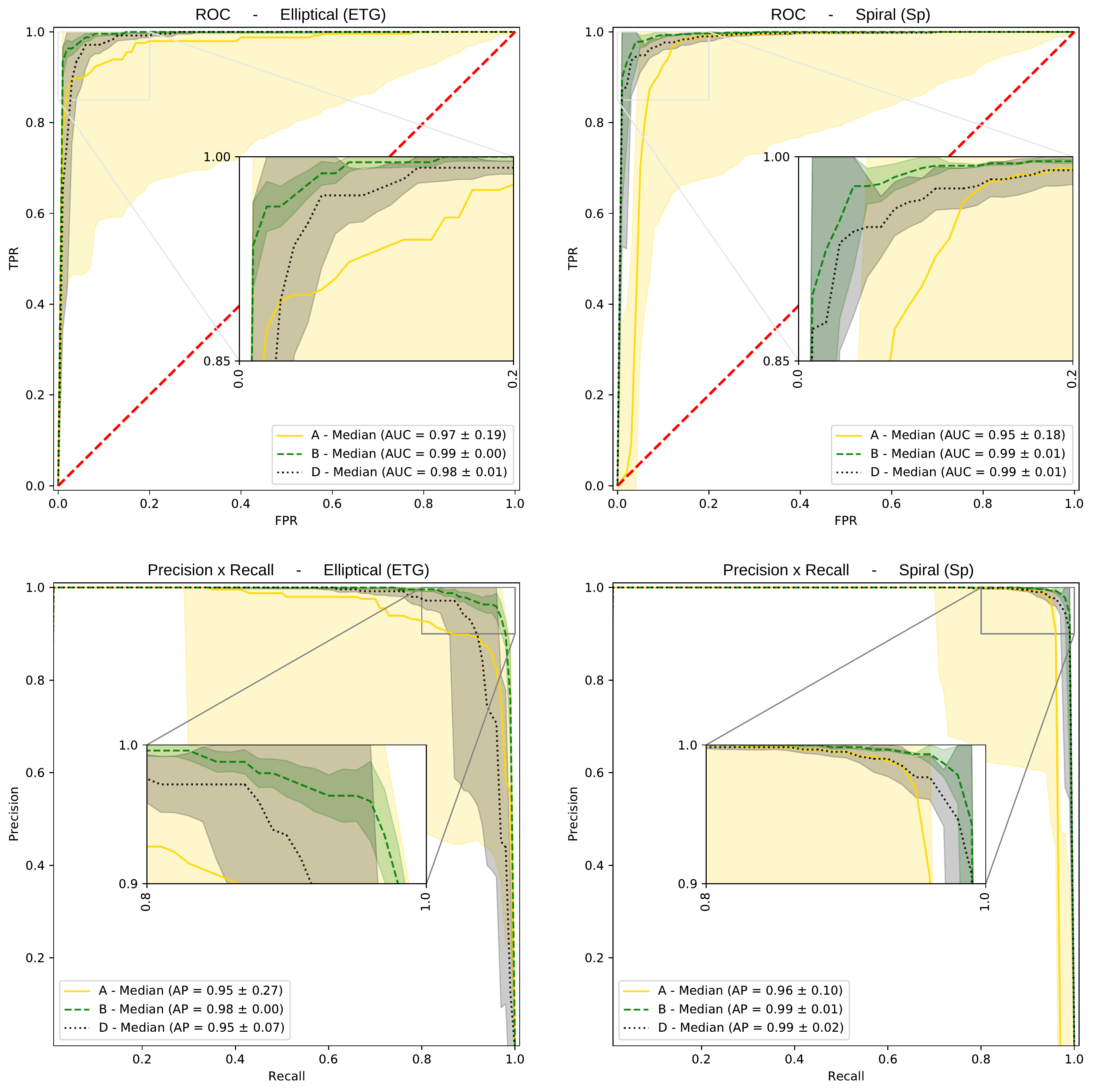}
    \caption{
    \label{fig:MC_EfficientNetB2_2}
    Results for A and D with the 3-, 8-band models, respectively. Model B used pre-trained ImageNet weights initialization with $3$ bands. The shaded regions represents the standard deviation in the folds. The upper panel presents the Receiver Operating Characteristic curve (ROC) where TPR is the True Positive rate (completeness or recall) and FPR is the false positive rate. The lower panels presents the Precision (purity) and Recall (completeness).
    }
\end{figure*}

It is worth noticing that all models presented here have a high performance in terms of AUC, with median AUC $>~0.98$. In Figures \ref{fig:MC_EfficientNetB2_1} and \ref{fig:MC_EfficientNetB2_2} we present the ROC for models A to ETG. For ETG galaxies, we noticed that all models, except the one with $12$ bands, have similar performance, close to ${\rm AUC}=0.99$ with $\approx 0.01$ sigma. The model pre-trained with ImageNet is more stable in terms of AUC, and this model is the one that gets closer to the perfect point $(0,1)$. The ROC in the Sp case favours the aforementioned pre-trained model, but also the $5$- and $8$-band models have a similar performance. We see that the model with $3$ bands only has a higher error in AUC and tends to deviate from the $(0,1)$ point if compared to the others.

The figures of precision and recall present a more clear picture of the models. We still see that the model with $3$ bands only with no pre-training performs worst in the median curve and has a high average precision error compared to others. The $12$ band model presents no gain compared to $5$ or $8$ bands.  The models with $5$ broad bands and $8$ bands have similar results. However, the first has a very stable result on average precision for S galaxies. The most notable difference, except for the aforementioned model, is that the model with $3$ bands only seems to be more sensible to the trials between test and validation for ETG galaxies. The $5$-band model also gets closer to the perfect-classifier point $(0,1)$ than the $8$-broadband model for ETG galaxies. However, the results show that $8$- and $5$-band differences are marginal. We present a more detailed picture considering the deviations in the five $K$-folds for the pre-trained model, the result for our base model with $3$ bands only, and the $8$-band model in Figure \ref{fig:MC_EfficientNetB2_2}.

\begin{figure}
\centering
\includegraphics[width=0.4\textwidth]{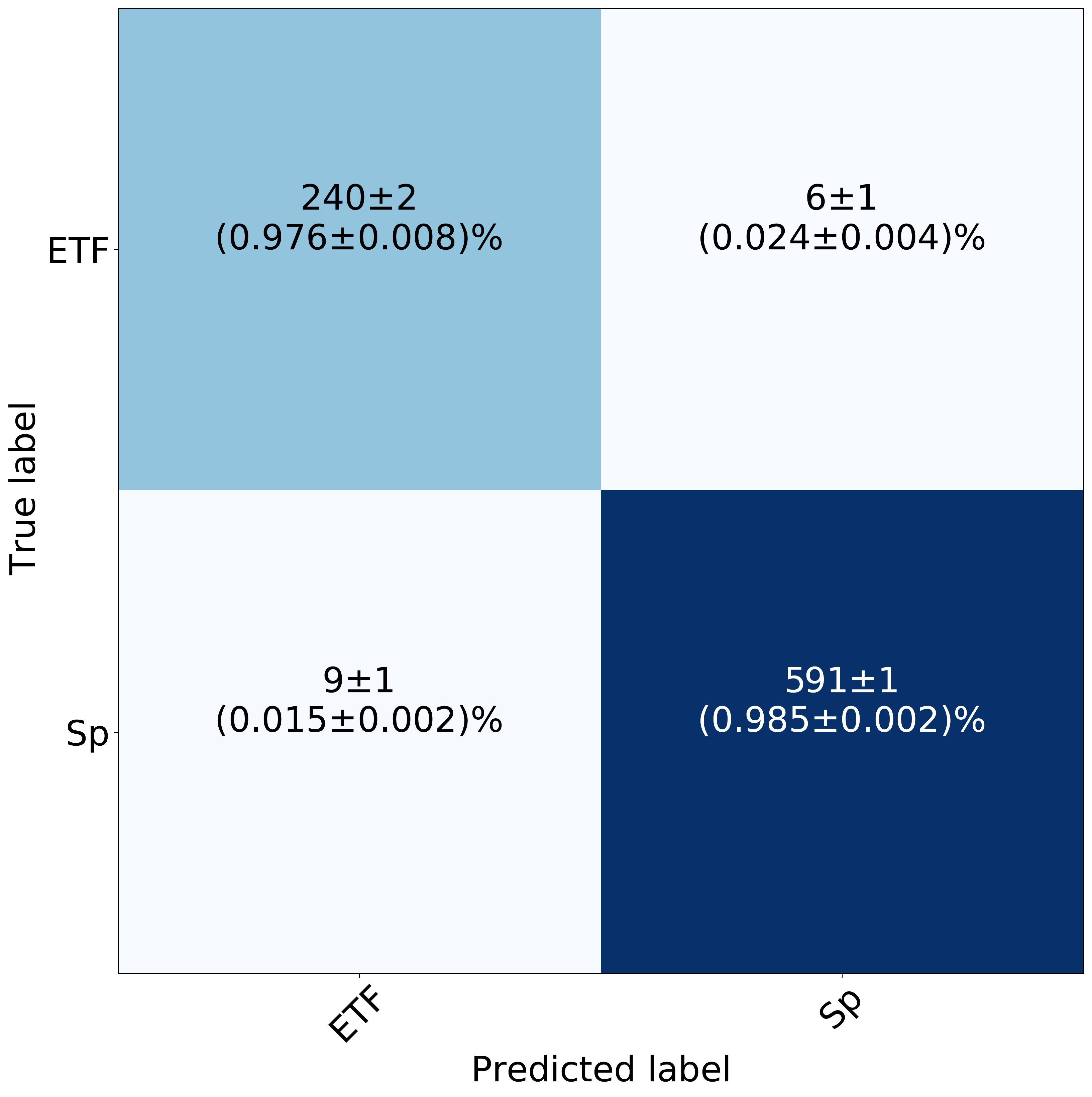}
    \caption{Confusion Matrix for the ETG-Sp classification on validation set, resulting from the best model provided by the deep learning.}
    \label{fig:MC_EfficientNetB2}
\end{figure}

Considering these results, we define the best threshold to be the one closest to the point $(0,1)$ in AUC. We present our results for the confusion matrix in Figure~\ref{fig:MC_EfficientNetB2} using the pre-trained network. This model consistently presented a high performance with high stability in both ROC and precision-recall curves and required fewer inputs than the other similar model with $8$- or $5$-bands. The confusion matrix represents the overall performance of true positives, true negatives, false positives and false negatives. Considering all the $K$-folds, we misclassify $15 \pm 2$ objects in our validation sample, representing less than $2\%$ (and therefore accuracy of $98\%$) for any  object with $r<17$ AB mag, in all redshift ranges considered.

\subsection{Application to ambiguous and Galaxy Zoo unlabelled dataset}

\begin{figure}
\centering
\includegraphics[width=0.95\linewidth]{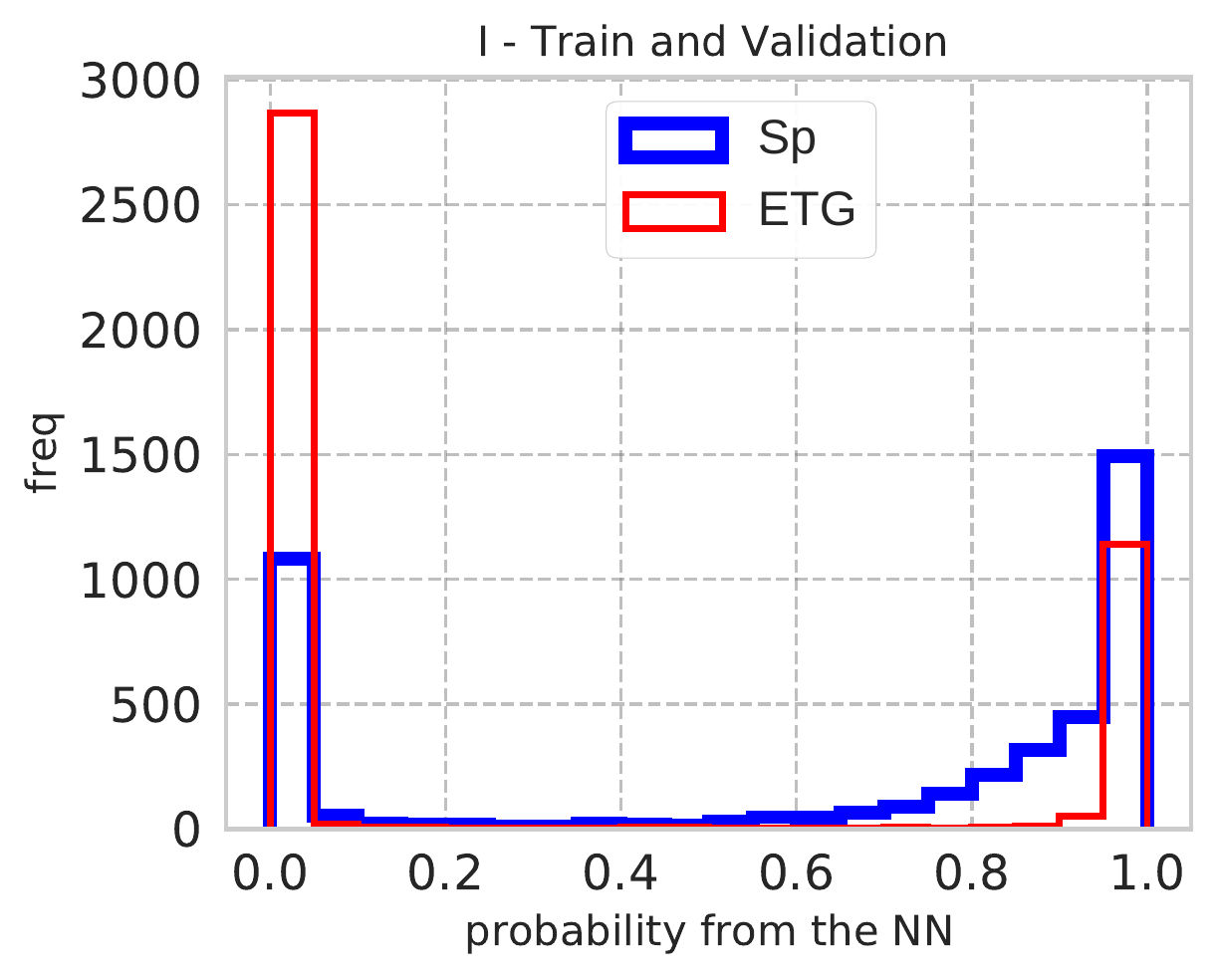}
\includegraphics[width=0.95\linewidth]{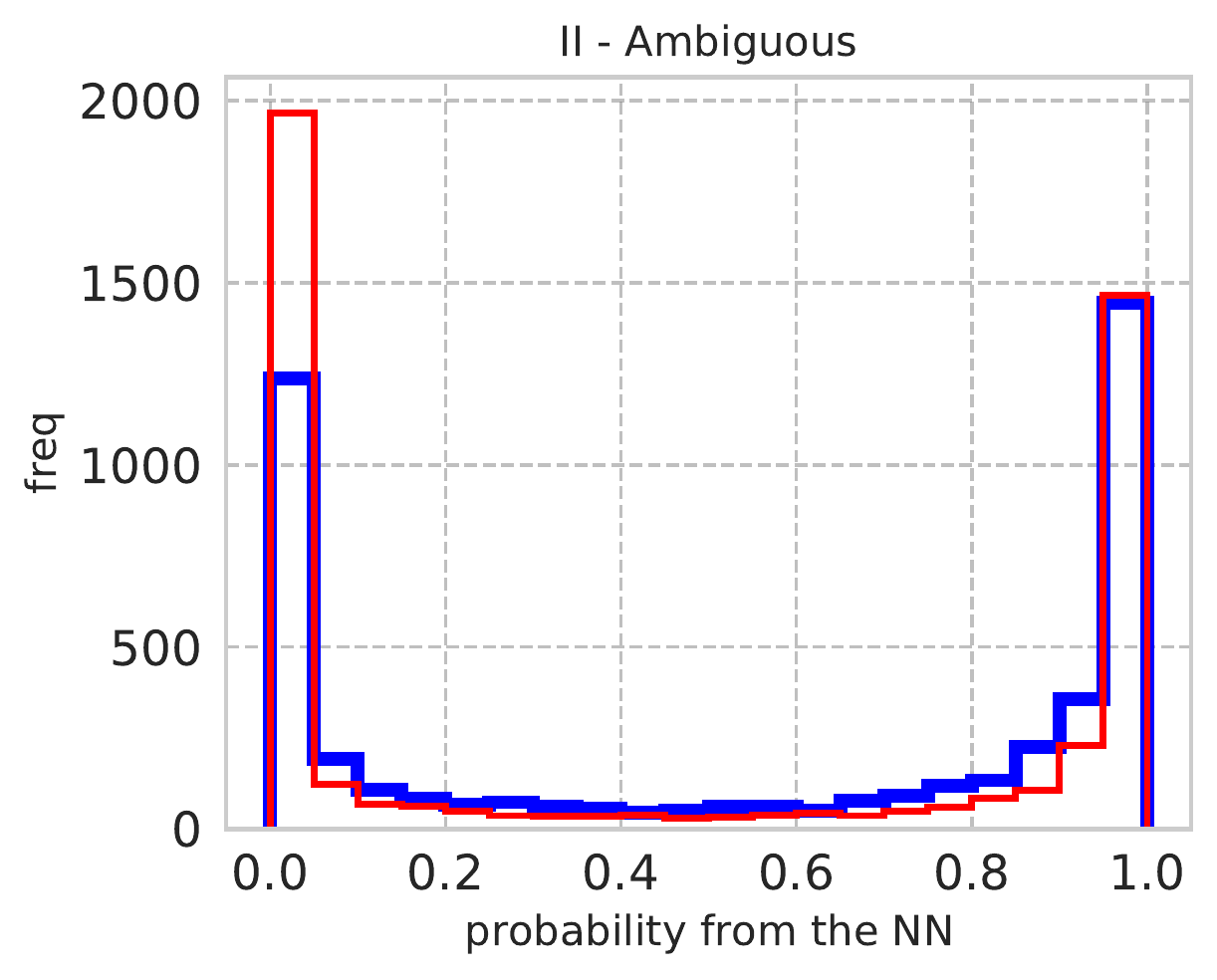}
\includegraphics[width=0.95\linewidth]{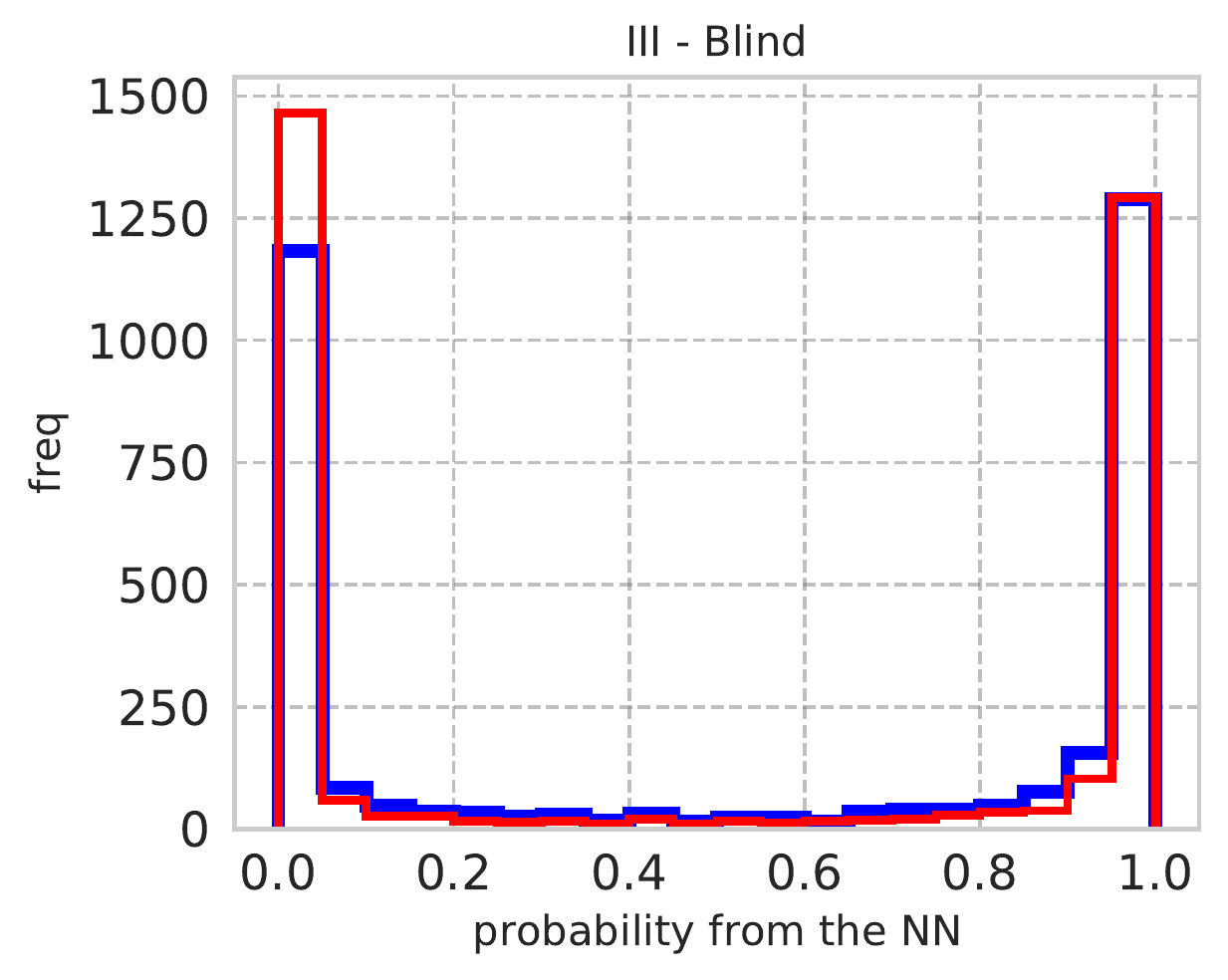}
    \caption{
    Distribution of probabilities (scores) for galaxies to be ETG or Sp provided by the Neural Network.
    \emph{Top} plot shows the probabilities obtained for the train and validation set (I). \emph{Middle}: probabilities for the ambiguous set (II) -- objects that have uncertain classification labels in GalaxyZoo.
    \emph{Bottom}: we present the probabilities for the blind set galaxies (III), where they have not classification labels in Table2 on GalaxyZoo.
    \label{fig:scores_probs}}
\end{figure}

We apply our best model, EfficientNetB2-like architecture using pre-trained weights from ImageNet in three bands to the other two samples.
In Figure~\ref{fig:scores_probs} we present a histogram that gives the score distributions for the three datasets: training and validation, ambiguous, and unclassified/blind samples. We emphasize that the distributions are qualitatively similar; the model presents high confidence in the non ETG probability end. 
The histogram showing the distribution of the scores for the ambiguous sample presents a more smooth distribution with respect to the blind and train and validation data-set, which might be an indicator that this sample presents a major challenge to the DNN. A detailed discussion of the sub-sample II can be found in the Appendix \ref{appendix}.

\subsection{Testing the S-PLUS ETG-Sp classification: comparison with other morphological classifications}
\label{subsec:results_morpho}

\begin{figure*}
\includegraphics[width=\textwidth]{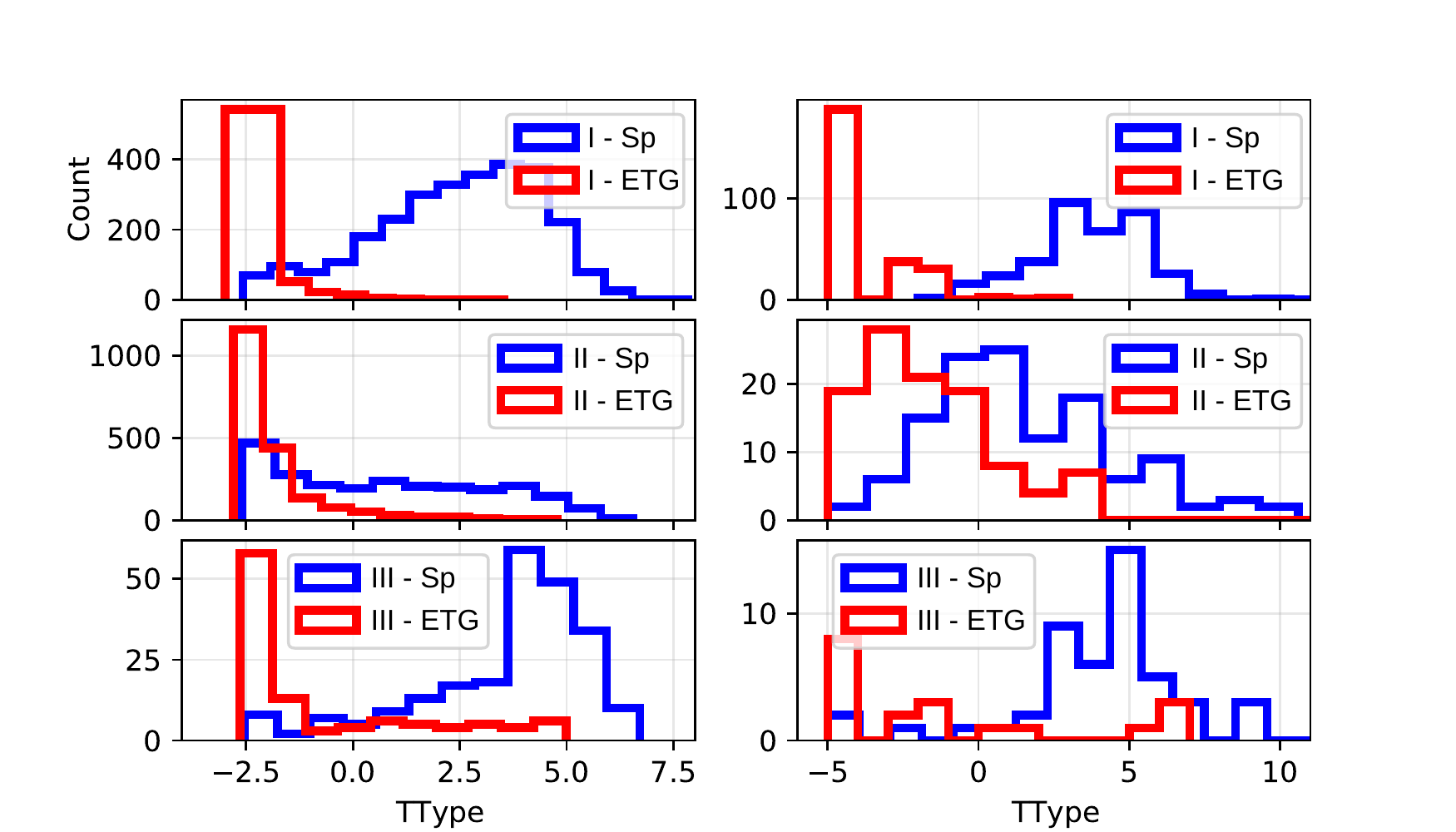}
\caption{{\it Left:} Histograms comparing the distribution of the ETG-Sp classification of this paper with respect to the T-Type morphological classification from \citet{Dominguez2018}, for the training-validation set {\it (top)}, the ambiguous set {\it (center)} and the blind set {\it (bottom)}.  {\it Right:} Histograms comparing the distribution of the ETG-Sp classification of this paper with respect to the T-Type visual classification from \citet{Nair2010}, for the training-validation set {\it (top)},  the ambiguous set {\it (center)} and the blind set {\it (bottom)}. In general, S0s have $-3 \le TType \simeq 0$.}
\label{fig:compDeep}
\end{figure*}

Figure \ref{fig:compDeep} shows the comparison between results from this work with other morphological classification of Stripe-82, based on DL (\citealp{Dominguez2018}; left column) and visual morphological classification (\citealp{Nair2010}; right column). It is worth noticing that \citet{Dominguez2018} DL method was trained by using GZ2 classification and \citet{Nair2010} classification.
In Figure \ref{fig:compDeep}, the first row shows the results for the training-validation set, the second row those for the ambiguous set and the last row those for the blind set. In the left column, the histogram showing the distribution of the galaxies classified as ETG has a peak at T-Type = -2.5 and generally the distribution falls below T-Type < 0 for the three cases. Galaxies classified as Sp have a broader distribution which peaks around T-Type = 4.5 for the training-validation and blind sample. Yet some galaxies classified as Sp have a T-Type lower or around zero, due to the difficulties of discriminating the presence of spiral arms in faint and higher-redshift galaxies, as discussed in Section \ref{subsec:GZ1_training}. This effect is stronger in the ambiguous class, exemplifying how this group of galaxies, if taken into consideration, would lead to an incorrect Sp-to-ETG galaxy ratio in the local Universe, and were therefore excluded during the debiasing process \citep{bamford2009galaxy}. The right column of Figure \ref{fig:compDeep} shows the histograms comparing the distributions of the ETG-Sp classification of this paper with respect to the T-Type visual classification from \cite{Nair2010}, for the training-validation set, the blind and the ambiguous sets. Unfortunately, when matching \cite{Nair2010} and the catalogue presented in this work, the sub-sample of objects in common is not large ($\ge 100$ objects). Even so, there is a clear agreement between the two classifications, with the only exception of  T-Type = 0 (i.e., the S0 class), which is hard to classify since it is not explicitly considered in the classification procedure of this paper and it will be implemented in Lucatelli et al. (in prep).

\subsection{Testing the S-PLUS E-S classification: morphological parameters of the newly classified galaxies}
\label{subsec:characterization}

\begin{figure}
\centering
\includegraphics[width=0.49\textwidth]{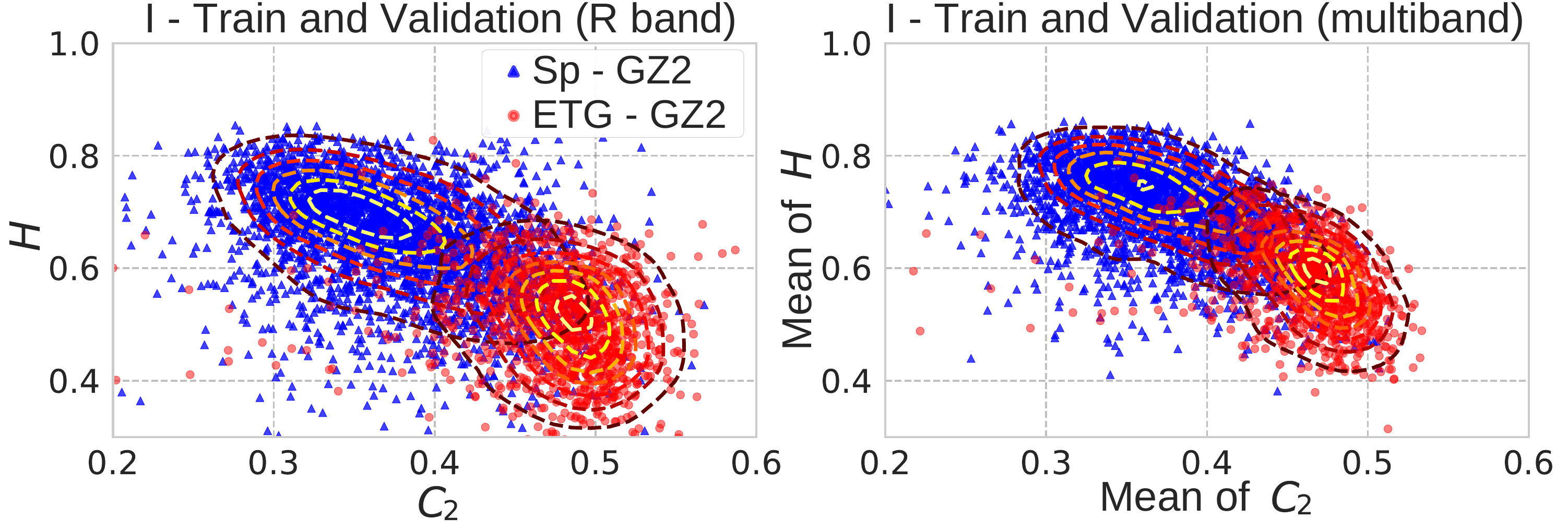}
\includegraphics[width=0.49\textwidth]{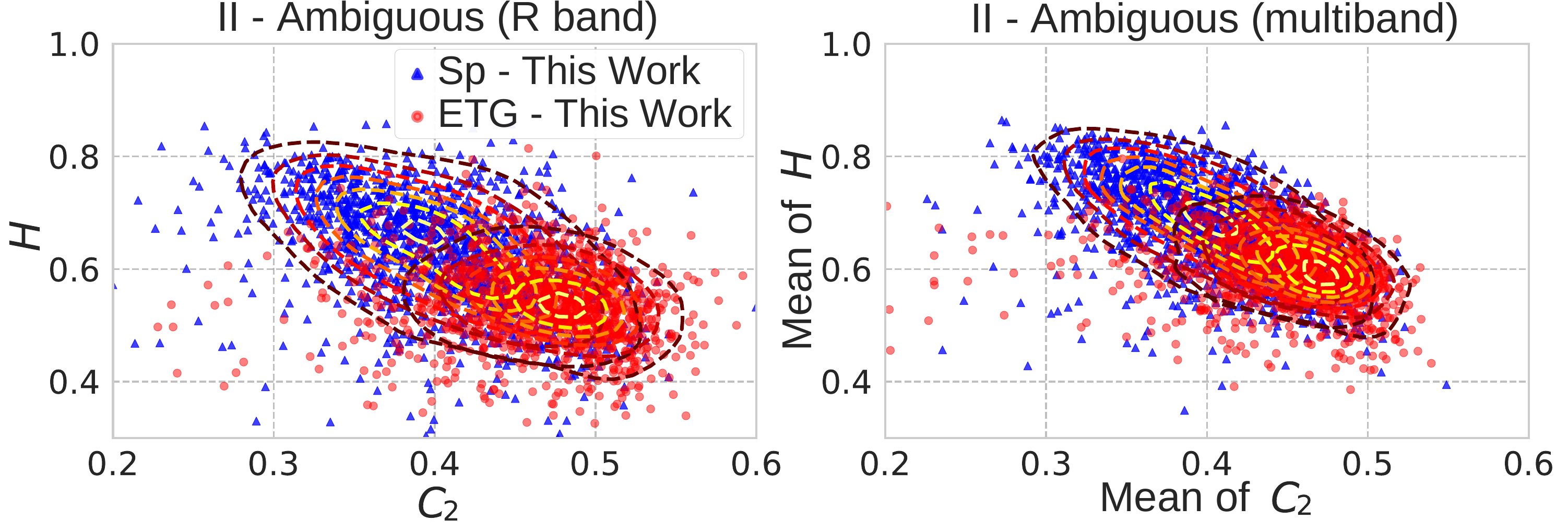}
\includegraphics[width=0.49\textwidth]{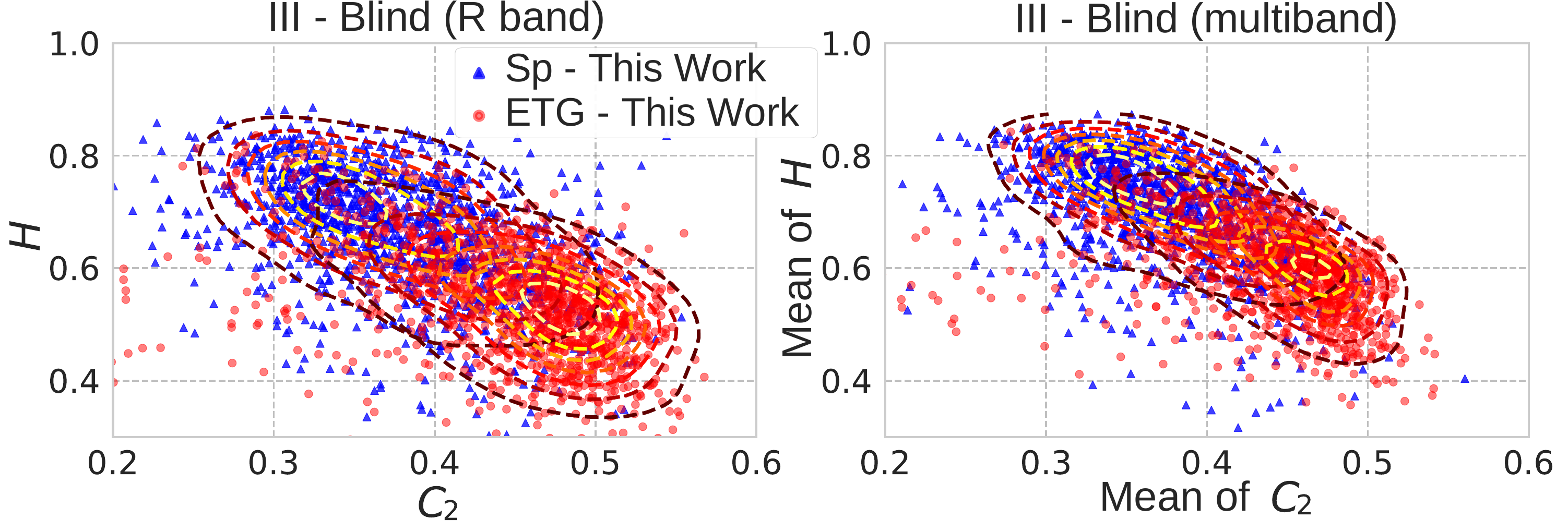}
    \caption{
    \label{fig:scatter_C2_H_S_E_all_sample}
    $C_2$-$H$ diagrams for our three sub-samples (I,II and III). 
   {\it Left:} Measurements from \textsc{mfmtk} performed in the $r$-band. 
   {\it Right:} mean value of $C_2$ and $H$ are computed for each galaxy using the u, g, F515, r, F660, I, F861 and z bands. The plots are colour coded according to the galaxy classification. 
   The two classes are clearly separated in this plane. Since the ETG-Sp classification is trained using GZ1 (i.e. visual classification based on the shape of the object) and the morphometric parameters also measure the monochromatic light distribution, this plot constitutes a consistency check: the two methods recover similarly the form of the objects, without taking into consideration the galaxy colour or stellar population.
   When considering the mean of $C_2$ and $H$ across different filters, there is less scatter in the distributions of morphometric indices.
   }
\end{figure}

In order to evaluate how the morphometric features would be distributed in the sample, we obtained measurements of $C$ and $H$. We compare these features to what we would expect for Sp and ETG galaxies, given what is already known in the literature. The determinations of these quantities for the sub-samples in Table \ref{tab:sample_description} are compared in 
Figure \ref{fig:scatter_C2_H_S_E_all_sample}. This Figure shows the $C_2$-$H$ morphometric diagrams for the three sub-samples (I,II and III), and the plots are colour coded according to the galaxy classification. All left plots refer to measurements only in the $r$-band and all plots on the right are made taking the mean of $C$ and $H$ in different filters.
The top of Figure \ref{fig:scatter_C2_H_S_E_all_sample} indicates the distribution of $C$ and $H$ for the training and validation samples, while the middle and bottom of Figure \ref{fig:scatter_C2_H_S_E_all_sample} refers to the same analysis but for sub-samples II and III, respectively. In these plots, the ETG and Sp classifications are the results of this work.

We note that, given that the ETG-Sp classification of this paper is trained using GZ1, i.e. visual classification based on the shapes of the objects (with or without spiral arms) and the morphometric parameters also measure the monochromatic light distribution, the plots in Figure \ref{fig:scatter_C2_H_S_E_all_sample} turns into a consistency check. The two methods, even if based in different techniques (classifications done by humans vs machines) recover the form of the objects, without taking into consideration the colour or stellar population of the galaxy.

The sub-sample measurements show a clear distinction between classes ETG and Sp, i.e., the centres of each contour lines are well separated. On the other hand, when looking to the ambiguous sub-sample, in the middle of Figure \ref{fig:scatter_C2_H_S_E_all_sample}, the centres of the two distributions are not well separated. This reflects the fact that these objects are classified as uncertain, as it is clearly seen in the relation between $C$ and $H$. Finally, for sub-sample III, at the bottom of Figure \ref{fig:scatter_C2_H_S_E_all_sample}, the centre of ETG and Sp distributions are better separated in relation to the sub-sample II, nevertheless, with a larger scatter than the classification from GZ1 (sub-set I at the top). 

Subsequently, when considering the same comparisons but using the mean of $C_2$ and $H$ in different filters (instead of using the $r$-band), we have a reduced scatter in the data. Still, the discrimination between `ETG' and `Sp' in sub-sample II is worst than for the others. 
The general conclusion is that the shape of the $C_2-H$ morphometric diagrams are similar to what is already shown in the literature \citep{Ferrari_2015, conselice2014} and when we use the classification of the deep-learning approach adopted here,  validating the results obtained in this work.

\begin{figure*}
	\centering
	\includegraphics[width=0.31\linewidth]{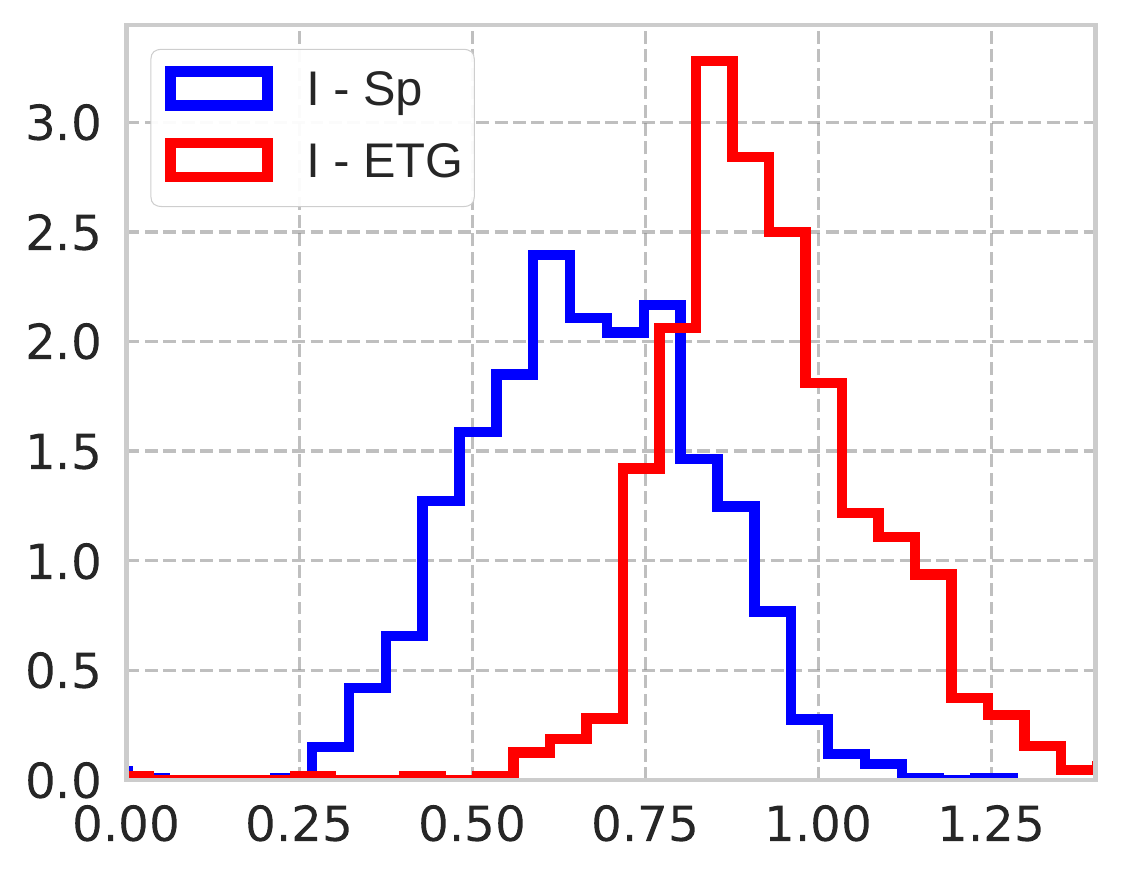}
	\includegraphics[width=0.327\linewidth]{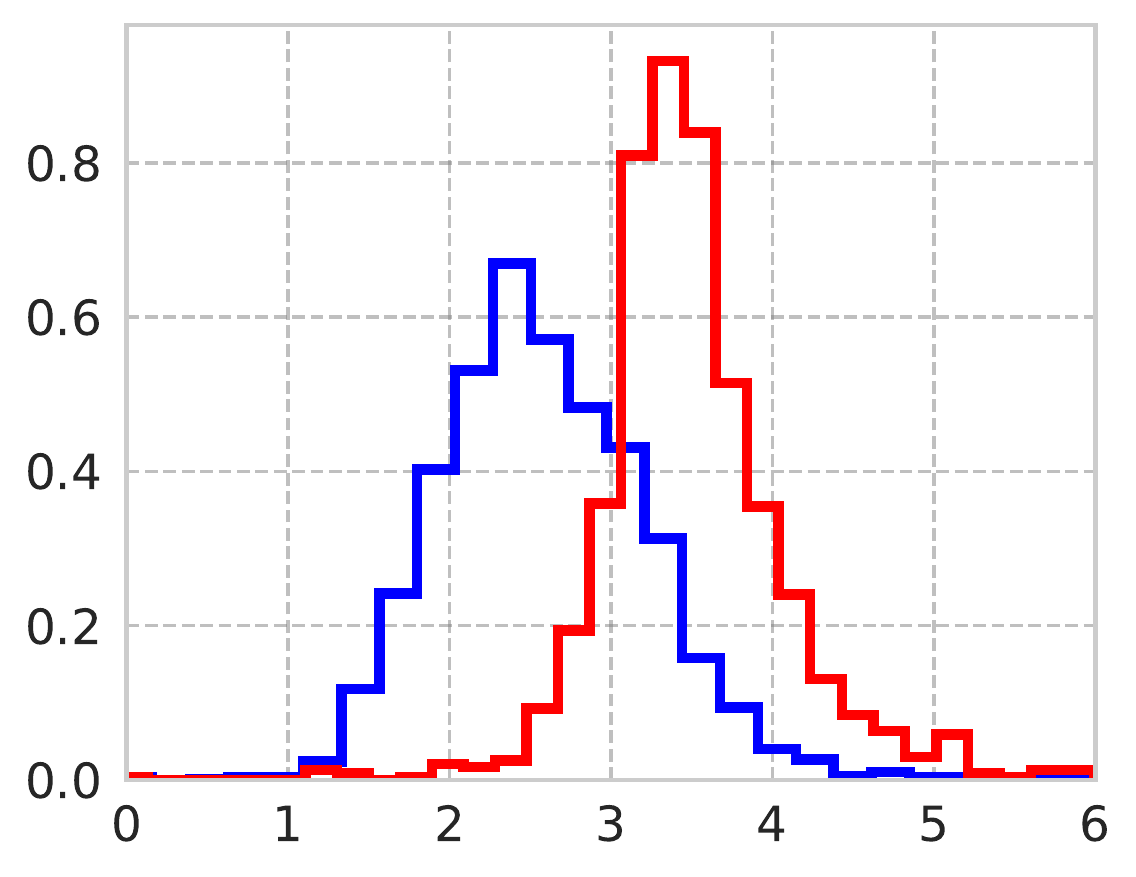}
	\includegraphics[width=0.338\linewidth]{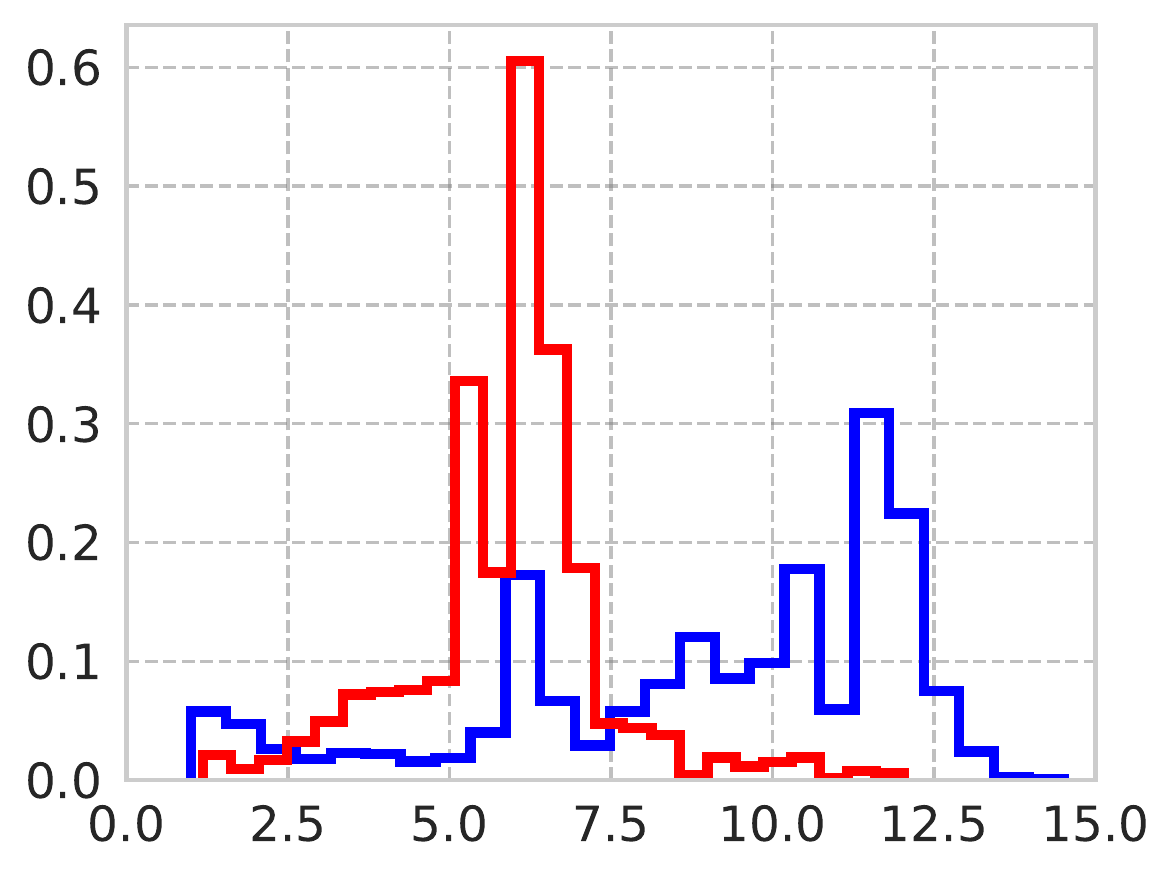}

	\includegraphics[width=0.31\linewidth]{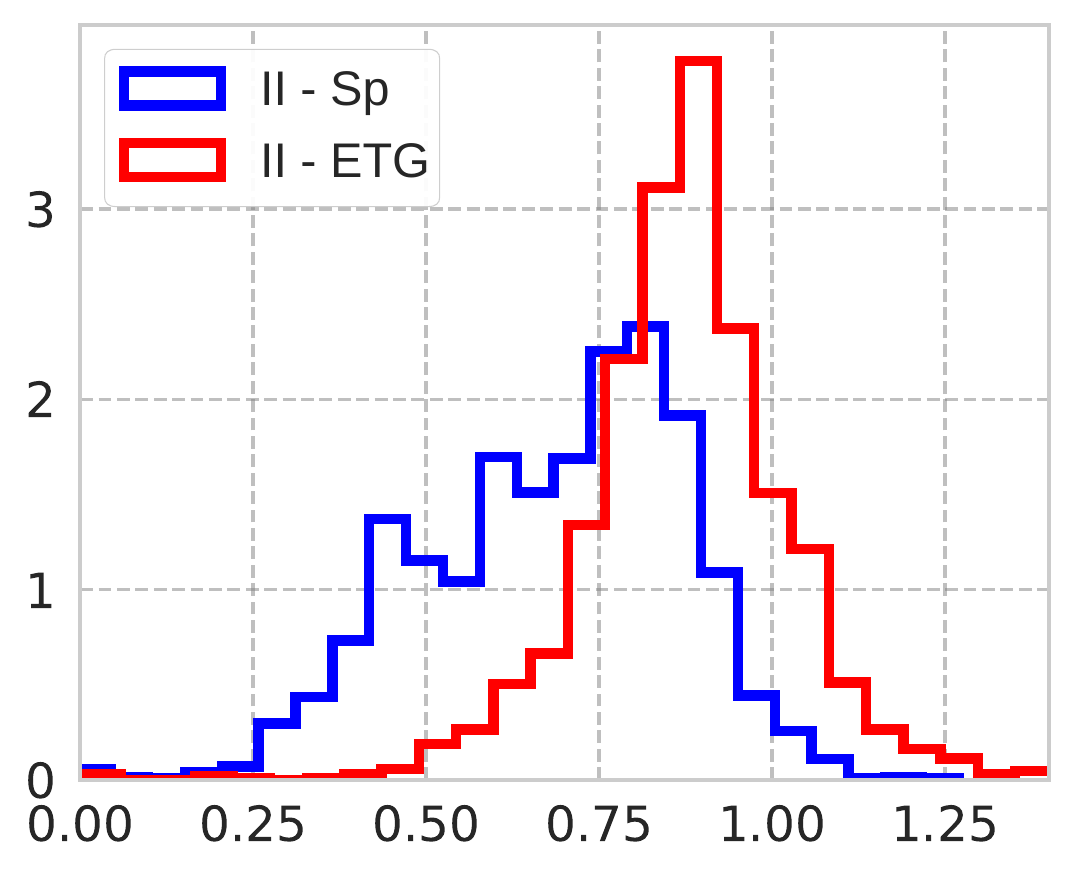}
	\includegraphics[width=0.327\linewidth]{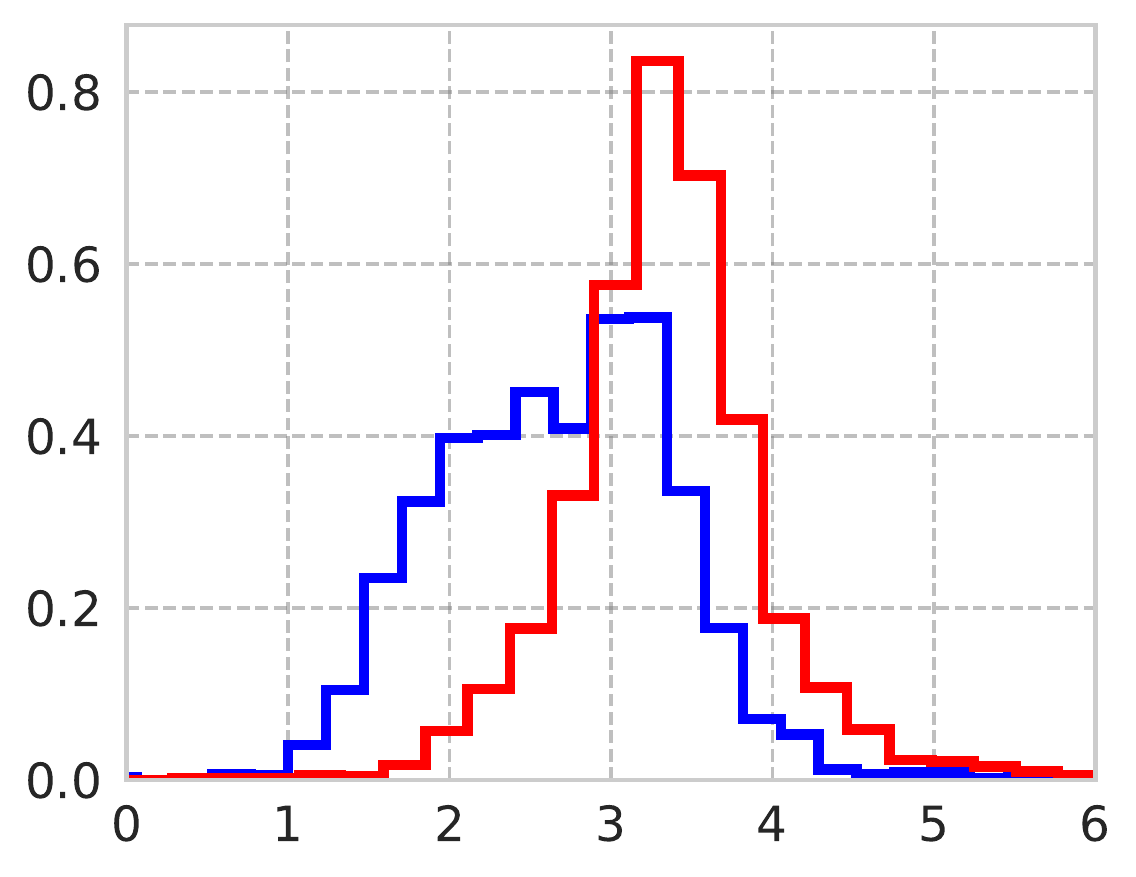}
	\includegraphics[width=0.338\linewidth]{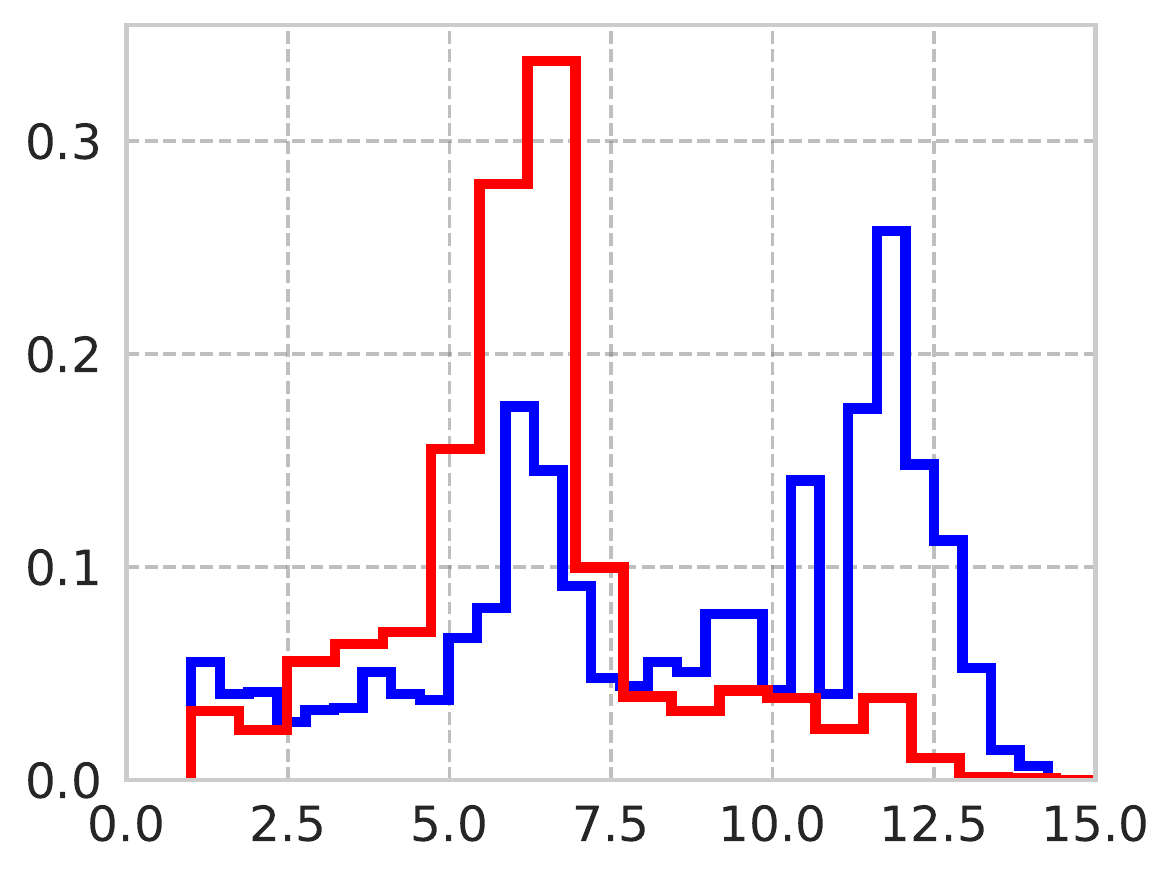}

	\includegraphics[width=0.323\linewidth]{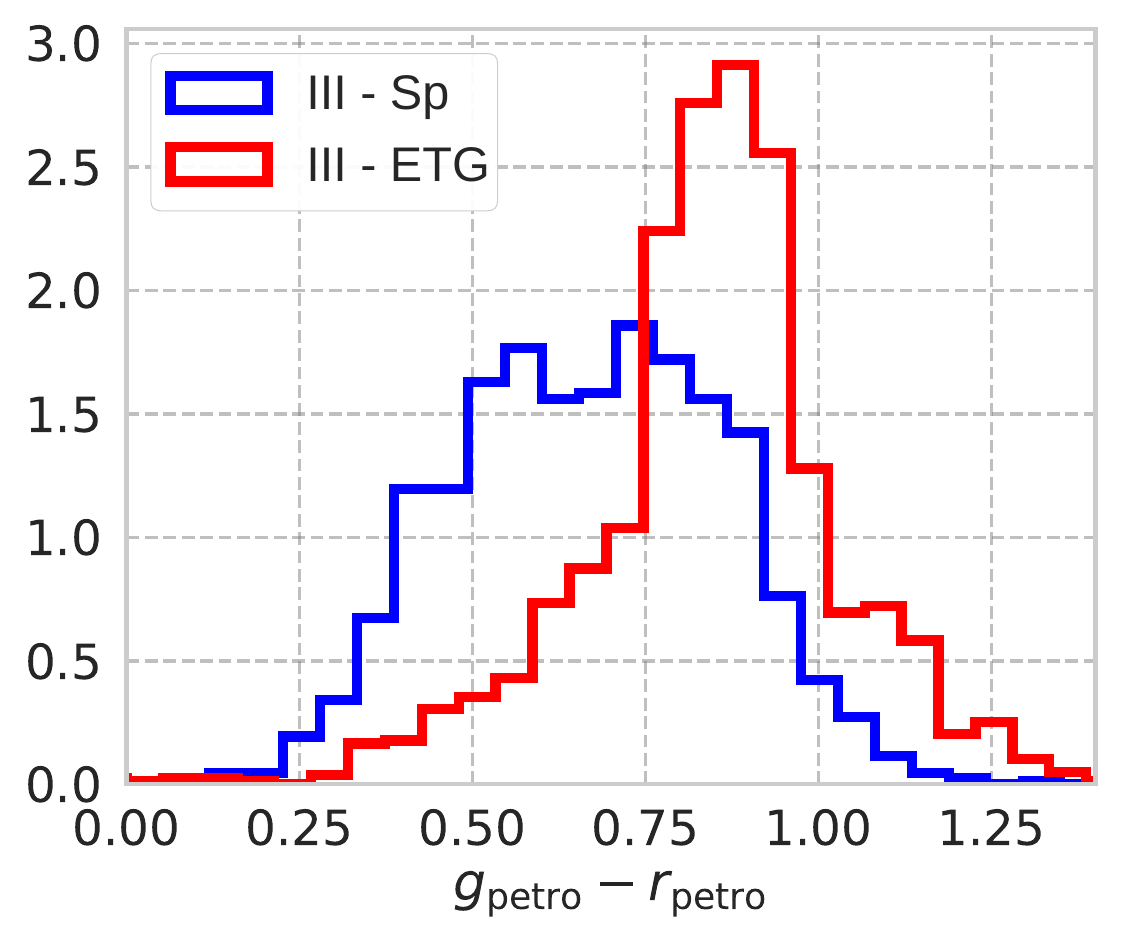}
	\includegraphics[width=0.325\linewidth]{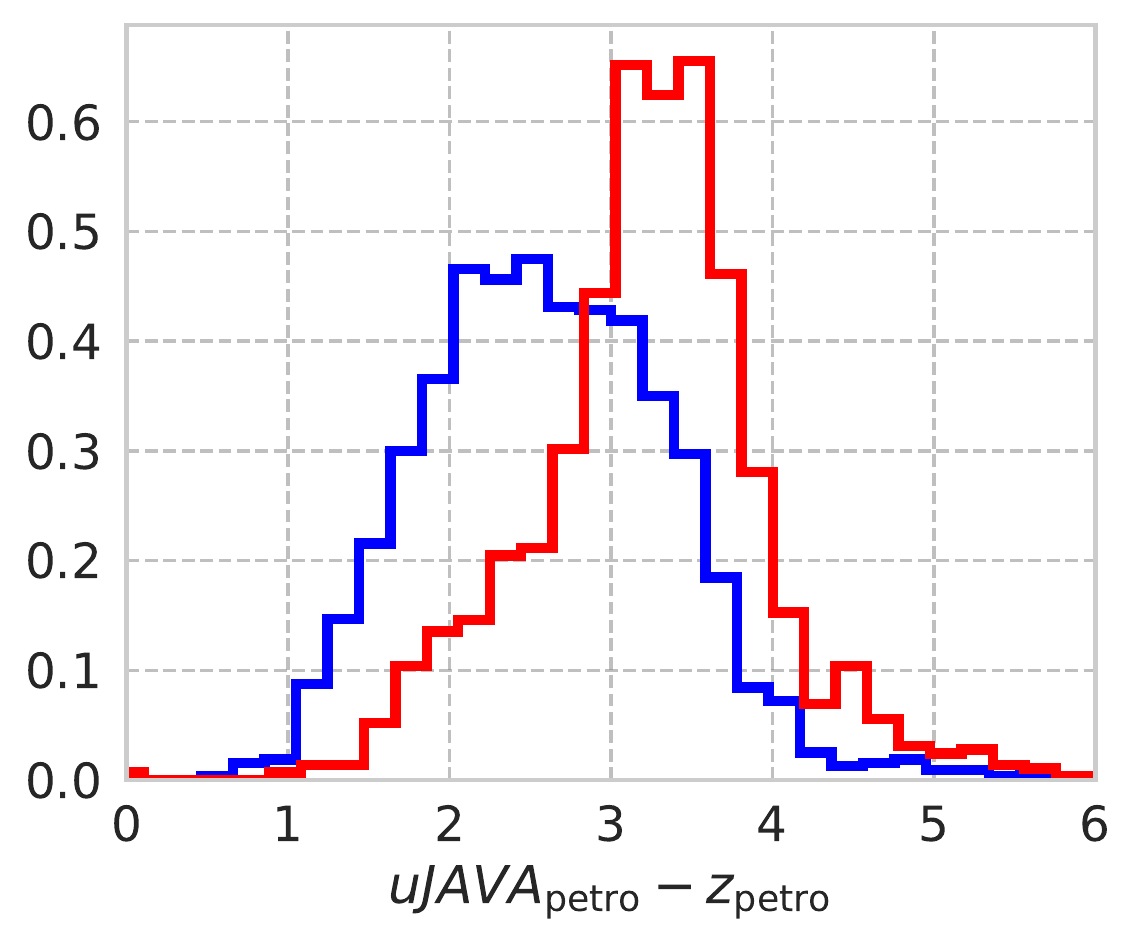}
	\includegraphics[width=0.333\linewidth]{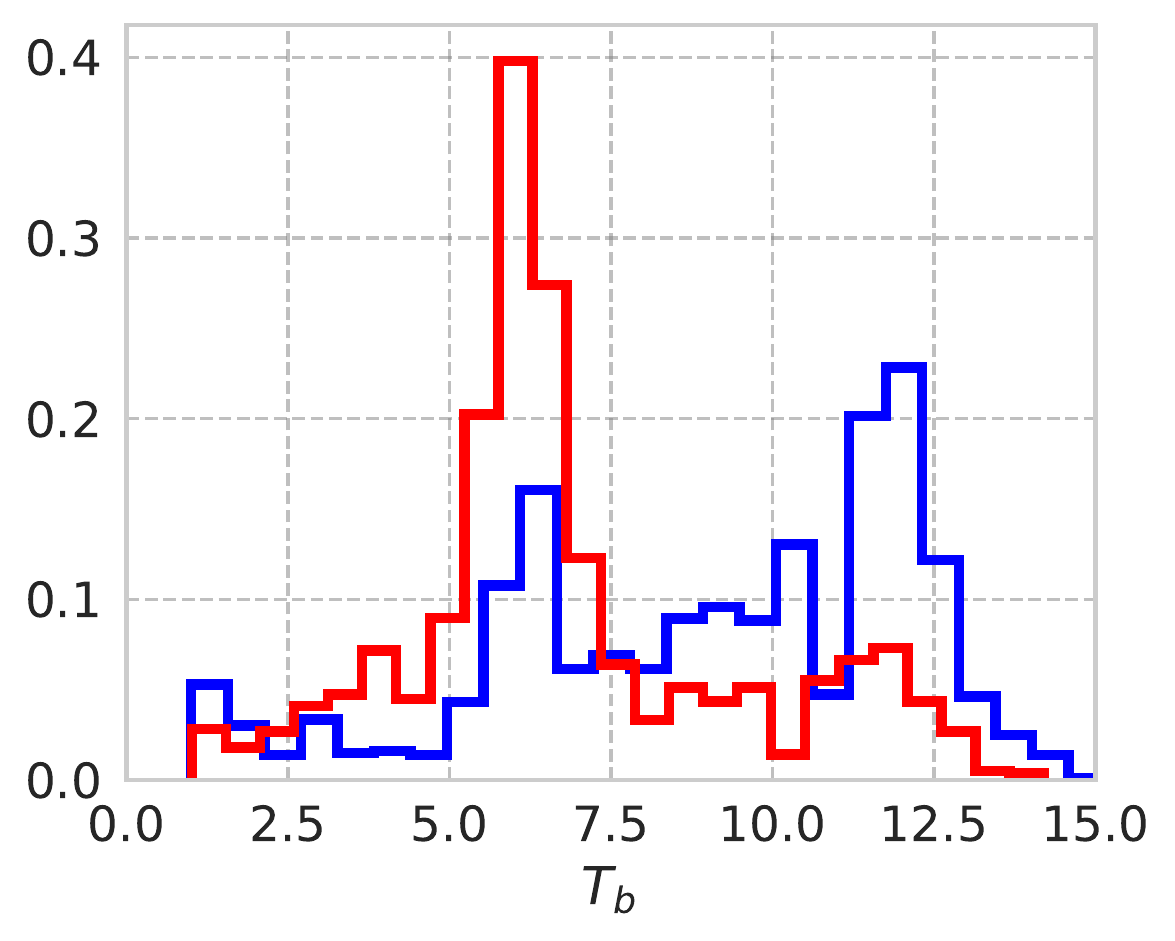}
	\caption{
	Histogram of the $g-r$ (left), $u$JAVA$-z$ (centre), and bayesian spectral type (tb; right) parameters for the ETG (red) and Sp (blue) galaxies in the training-validation set (I) ({top}), the ambiguous set (II) ({middle}) and the blind set (III) ({bottom}).
	}
	\label{fig:galaxies_properties}
\end{figure*}

Finally, the left and central panels of Figure \ref{fig:galaxies_properties} show the $(g-r)$ and $(u-z)$ normalized colour distributions of the galaxies, colour coded according to the galaxy classification, for the three sub-samples (I, II and III). The two galaxy populations, ETG and Sp, present the expected colour bimodality, with classified ETG's being redder than Sp's. 
Yet, there is a region of overlap where blue ETGs and red Sps fall, as already pointed out by the works of \citet{bamford2009galaxy, Schawinsky2009, Masters2010}. 
The third column shows the distribution of the Bayesian spectral type (Tb) parameters, as obtained by \citet{Molino2020}. Briefly speaking, to derive a photometric spectral measure (photo-z) from the galaxies' SED it is necessary to fit, together with the redshift value, the most probable spectral type of the galaxy. The redshift is in fact determined from the shift of the observed SED with respect to galaxy templates used in the fitting routine. In this case, the results are obtained using BPZ \citep{Ciccio} and are presented in \citet{Molino2020}.
The distribution of the Tb parameter shows that when the whole galaxy SED  is taken into account, the complexity of the relation between galaxy morphology and the stellar population, already visible in overlap areas the colour distribution of ETG-Sp galaxies, is even more complex: quenching effects, as well as the influence of star formation bursts become detectable using a full SED fitting or some specific narrow bands.
This plot also reveals the potential of S-PLUS survey in studying galaxy evolution and formation.

\begin{figure*}
\centering
\includegraphics[width=0.33\linewidth]{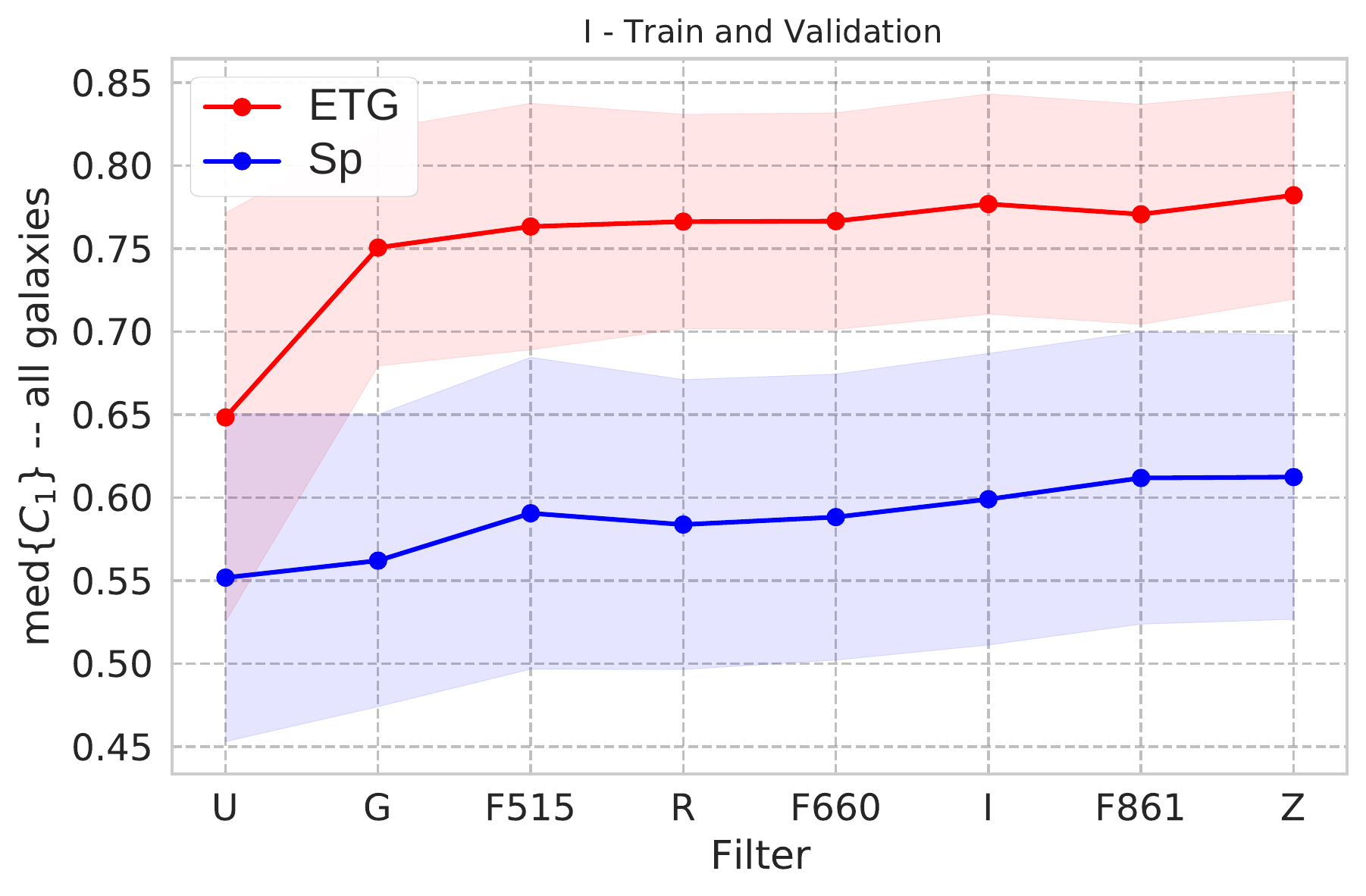}
\includegraphics[width=0.32\linewidth]{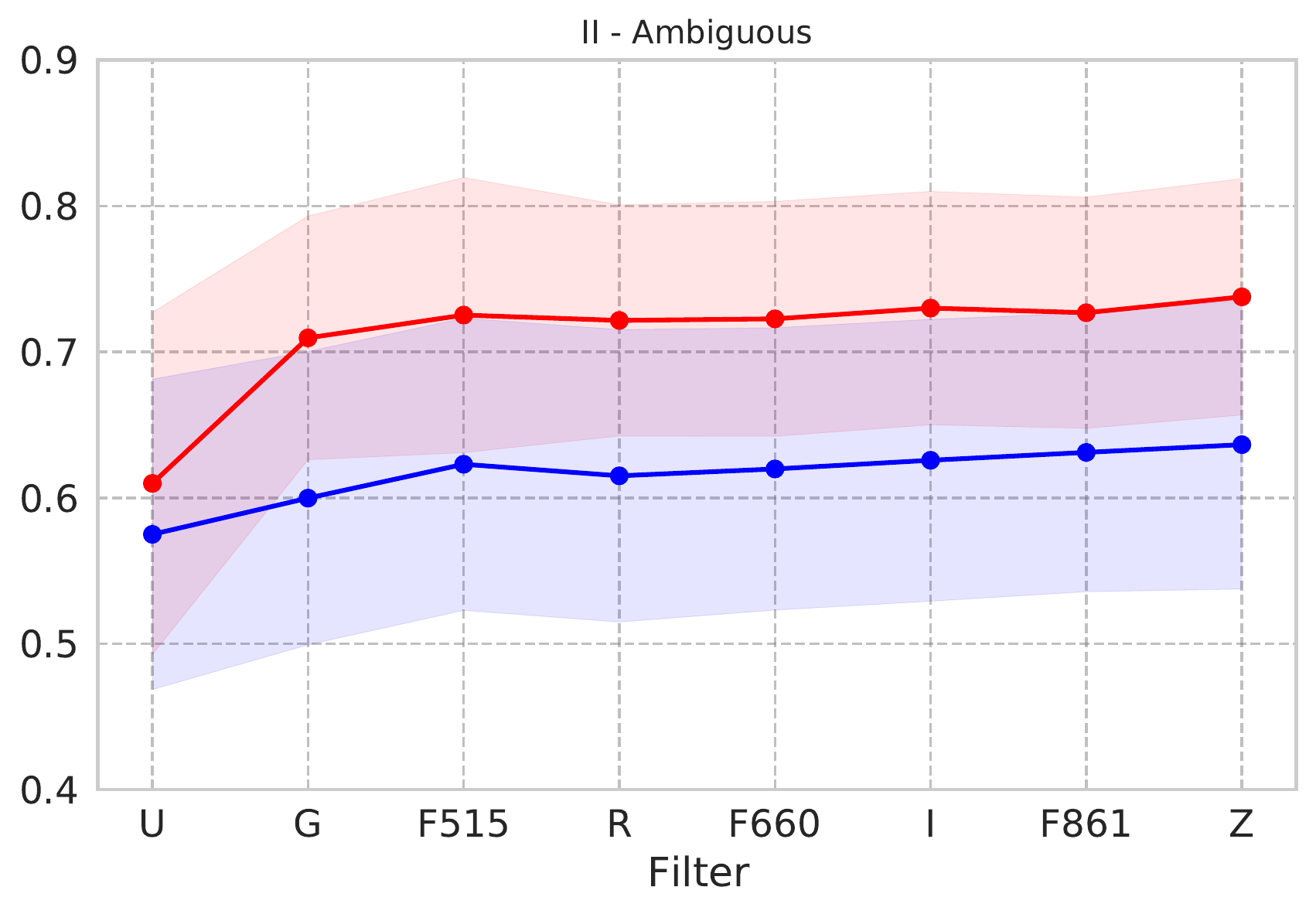}
\includegraphics[width=0.32\linewidth]{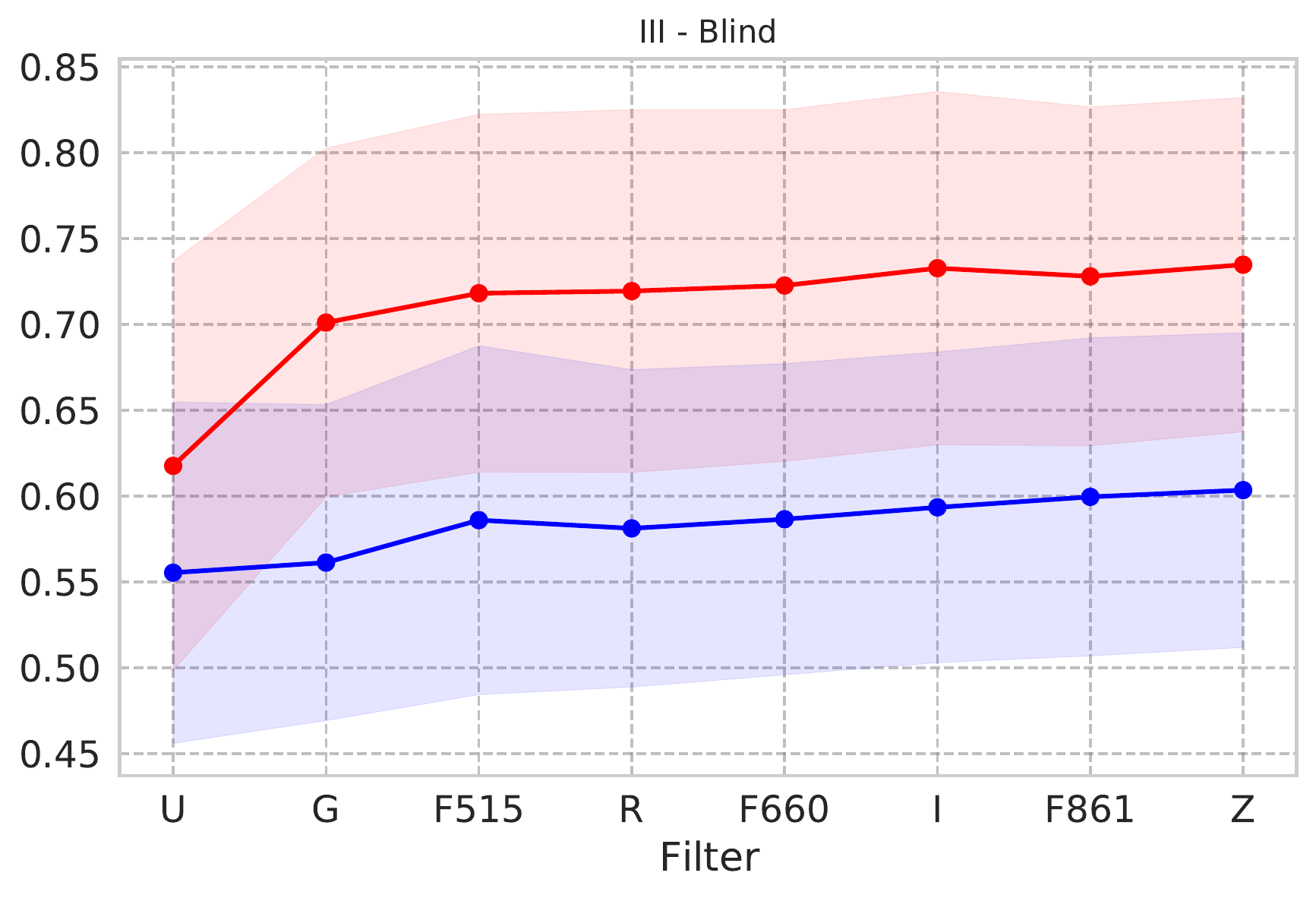}
\includegraphics[width=0.33\linewidth]{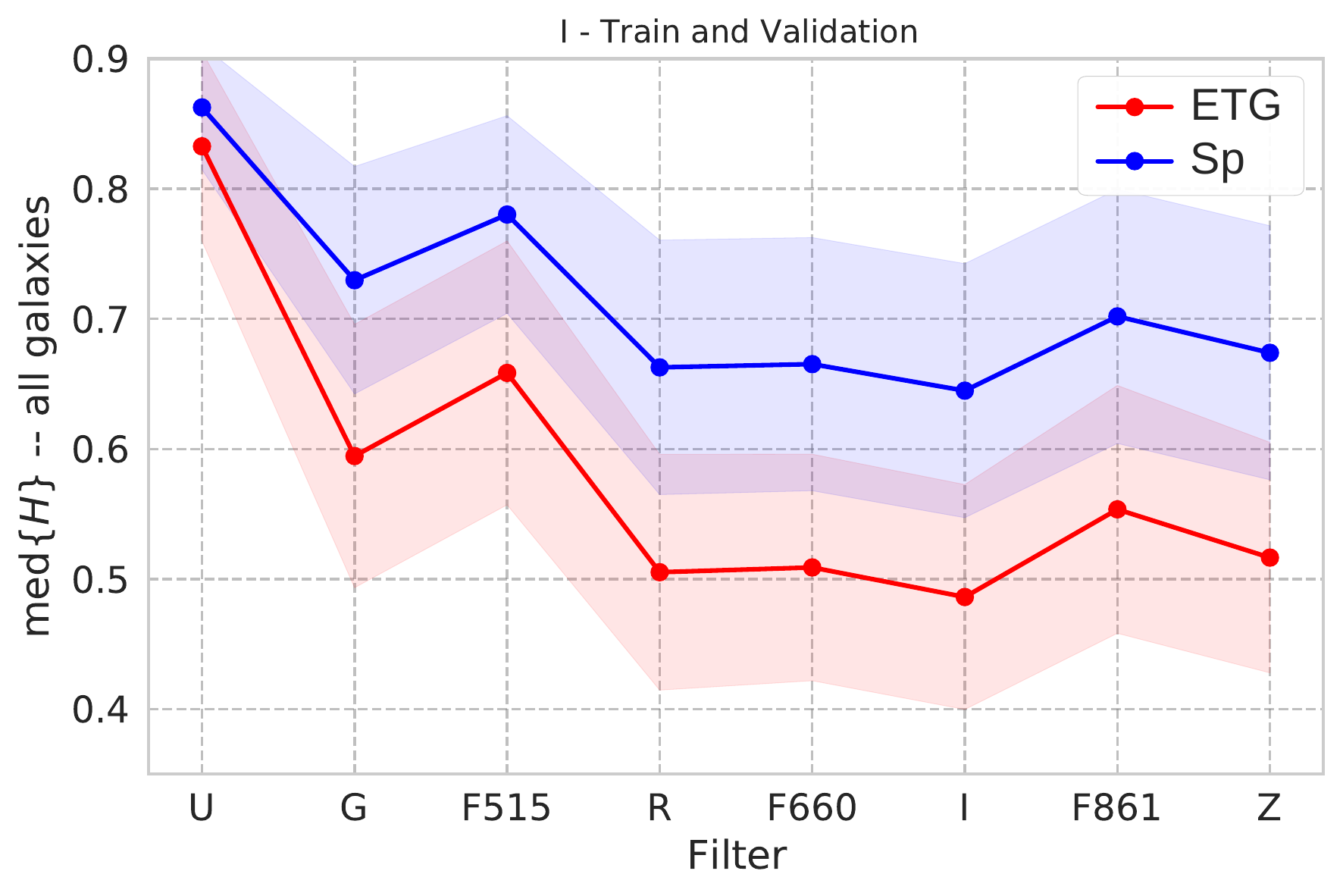}
\includegraphics[width=0.32\linewidth]{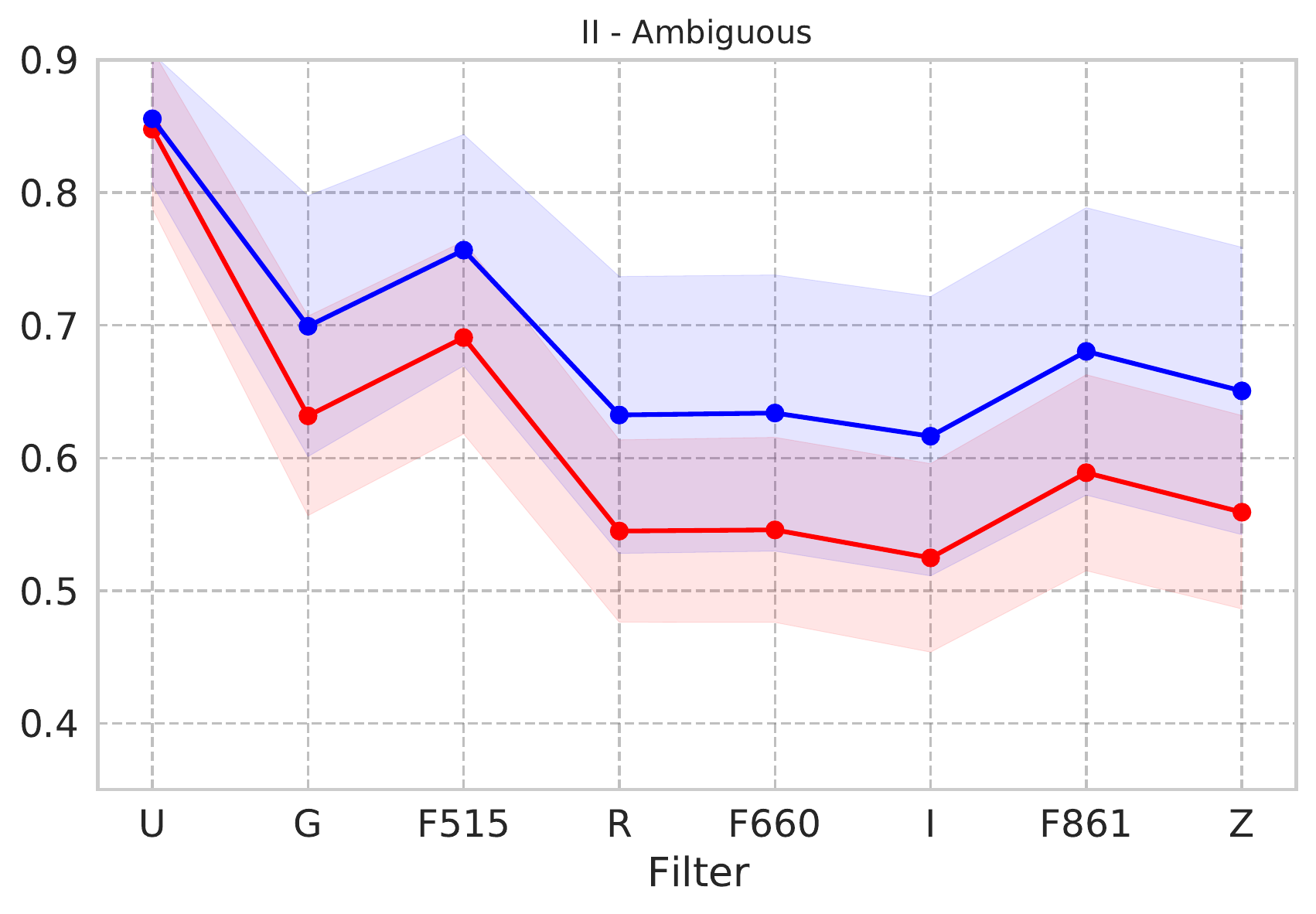}
\includegraphics[width=0.32\linewidth]{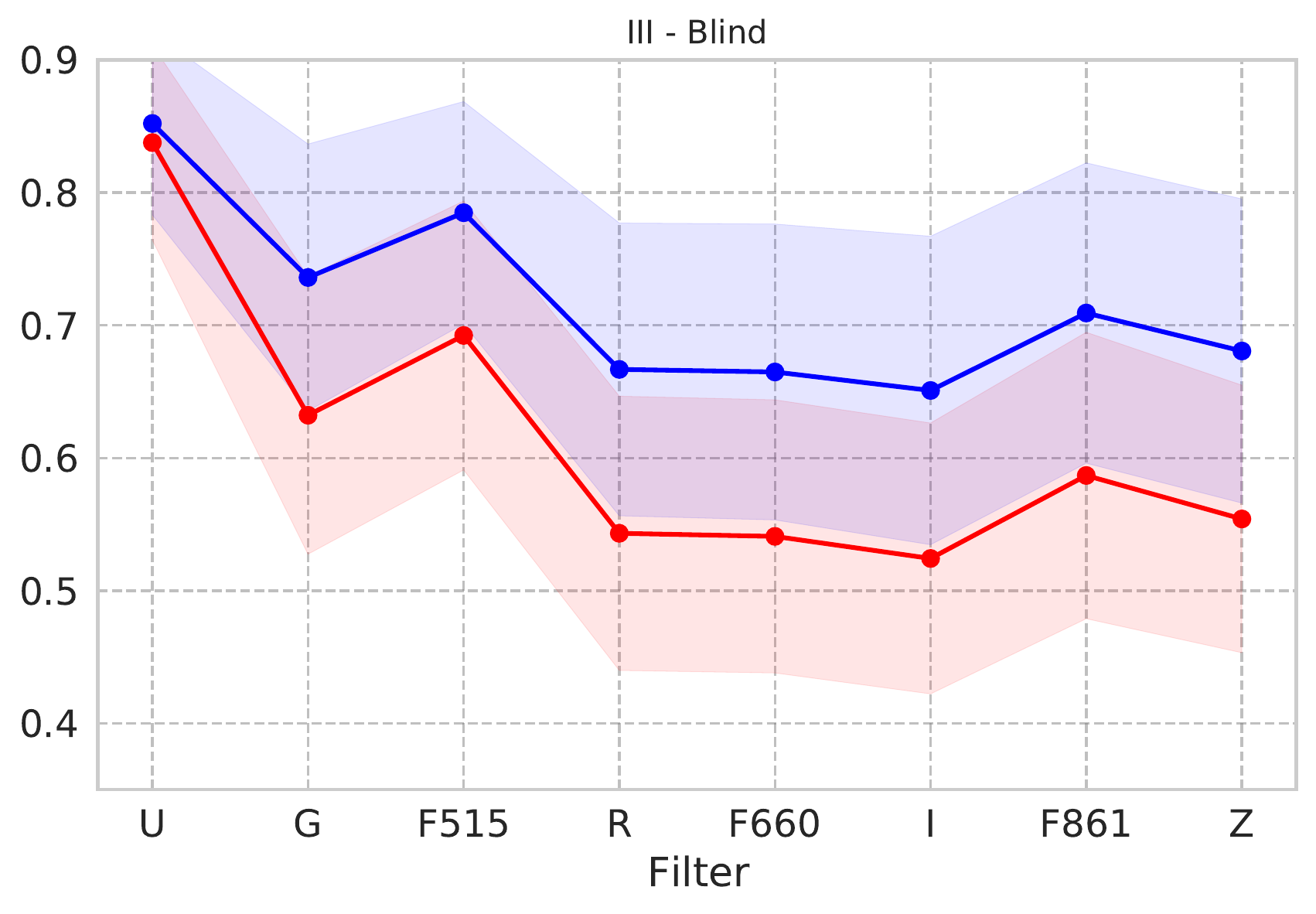}
    \caption{
    \label{fig:evo_C_H} 
    Evolution of morphometric measurements in different filters for ETG and Sp galaxies. Each dot represent the median value of the measure ($y$-axis) for all galaxies in a specific filter ($x$-axis). Therefore, we are observing how the median value of a given quantity changes globally for all galaxies as a function of filter. The shadowed errors are $\pm 1\sigma$.
    \emph{Top:} Measures for concentration $C_1$ in the train and validation set (left), ambiguous set (centre) and blind set (left). 
    \emph{Bottom:} The same as $C_1$ but for entropy $H$.
    }
\end{figure*}

\section{Discussion and concluding remarks}
\label{sec:discussion}

\subsection{Summary}
In this paper, we use the state-of-the-art of deep learning methods for image classification and present a model for predicting astrophysical features of nearby galaxies in Stripe-82 observed within S-PLUS. Specifically, we recover their morphology, assigning each object a probability of being a Sp or an ETG galaxy, using artificial intelligence. 
We also evaluate the use of $12$ bands presented in S-PLUS. We use different combinations of filters (broad and narrow) to check the effect of the inclusion of narrow bands in the analysis. In fact, morphological classification using DL algorithms has been already implemented in several works \citep{Dominguez2018, Barchi2020,tuccillo2016deep,khalifa2018deep,dieleman2015rotation,zhu2019galaxy,dai2018galaxy,Gupta2020,Vega-Ferrero2020}. In this work, we investigate the gain/loss of adding narrow-band images to perform the classification.  We also evaluate the use of weights extensively trained in other reference computer vision dataset to initialize the DNN. 
We deliver a new catalogue of galaxies morphologies, including a classification for galaxies considered ambiguous or unclassified in GZ1.  

\subsection{A New catalogue of Galaxy Morphology}
\label{sec:new_catatlogue}
The final catalogue we developed can be divided into three samples. The first was used as training and validation and it was already classified in GZ1 catalogues. Therefore, this sample acts as a quality control sample. This sample was crucial to develop working and high performance DNN models in S-PLUS. The model created from this set will pave the path towards the morphological classification in the whole S-PLUS footprint.

The second sample contains the galaxies considered ambiguous in GZ1. Our assessment using independent features to analyze the data suggests that these objects are indeed ambiguous in the parameter space of the   morphological features used ($C$ and $H$). However, we do see some degree of difference in the E and S groups classified by the current DL method. It is worth noticing that this group has more objects in the faint end than the control group.

The last group contains the objects that are not presented in  GZ1 catalogue, either because they lie at a redshift lower than $z<0.03$, or because they do not have redshift estimate required for the debiasing procedure, or because they did not reach the number of votes needed to be classified in GZ1, among other reasons. Thus, we present a new classification for these objects using S-PLUS data with competitive quality to the human/machine performance used in GZ1.

The three groups together contain all the S-PLUS DR1 galaxies in Stripe-82 with r$_{\rm petro}$ < 17, see Section \ref{sec:data} for more details.

\begin{figure*}
    \centering
    \includegraphics[width=0.8\linewidth]{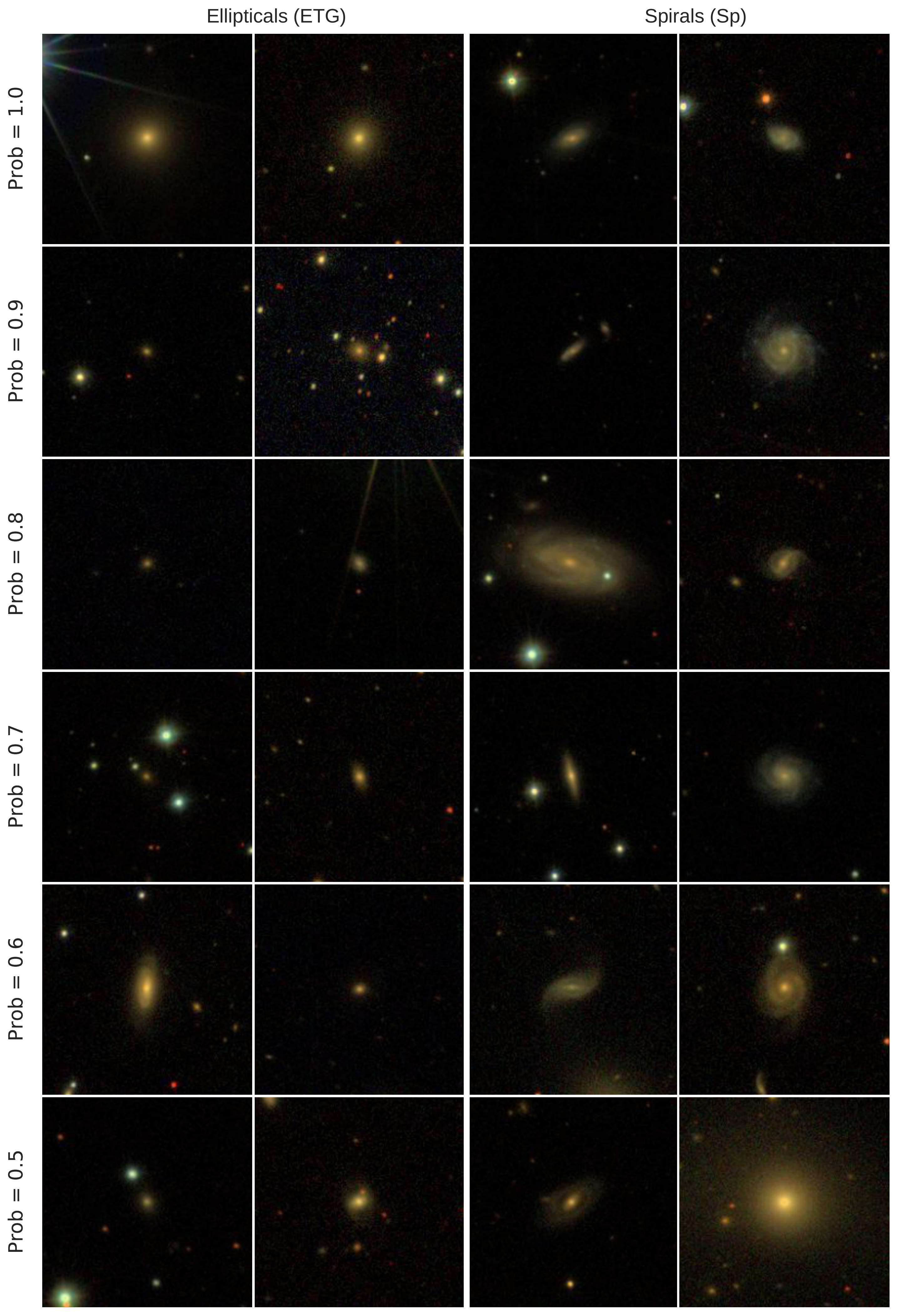}
    \caption{ From top to bottom: Example of galaxies images as classified in this work, for decreasing values (from 1 to 0.5) of probability of being spiral (first and second columns) and elliptical galaxies (third and fourth column). }
     \label{fig:galaxiestypes}
\end{figure*}

In Figure \ref{fig:galaxiestypes} we present a visual assessment of randomly chosen galaxies in the full catalogue in a certain confidence interval. It is clear that somehow we are biased toward the objects that have a higher number in the sample, i.e. as galaxies in top row: the images are quite shallow. Nevertheless  the deep learning algorithm is doing a great job in classifying them as Sp's or ETG's: we have more obvious cases with larger and more clearly visible (to the  human eye) spiral arms. Yet, since these types of galaxies are less present in the training sample, they get  slightly lower values of probability (0.9). When the probability gets lower than 0.7 some cases of confusion are present. It is  possible to see in Figure \ref{fig:scores_probs} that the number of cases with a prediction lower than 0.7 is scarce (in the case of the ETG galaxies, around 3 galaxies per bin of probability). The catalogue released within this contribution contains the probability of each class and also our classification using the same threshold imposed in the confusion matrix described in Section \ref{sec:model performance}.

 \subsection{The effect of using narrow bands}
 
 We compared 5 different models of DL: using 3 broad bands (with and without pre-training), 5 broad bands, 8 bands (5 broad bands and 3 narrow bands) and 12 bands. Note that this is the first time that automated multi-band galaxy classification is applied to more than 4 bands \citep{Vega-Ferrero2020}.
 Looking at Figure \ref{fig:MC_EfficientNetB2_2}, it is possible to see that the best model is the one with 3 bands, pre-trained with the ImageNet, highlighting the importance of pre-training in this type of analysis (note however that for the  moment pre-training is possible only for 3 bands as the ImageNet images contain three channels). In fact, when comparing the remaining four cases with no pre-training, it is clear that the 3-band case has the worst performance: as mentioned earlier, it was the only one that did not optimize the training loss in all the folds. Interestingly, the 5- and 8-band models have the highest precision for recall (among the models with no pre-training), suggesting that using all the 12 bands, including some low  S/N bluer narrow bands, might increase the confusion in the classification procedure, considering a DNN architecture with the same complexity.  Not surprisingly, in the 8-bands model, the three narrow bands used are F515, F660 and F861, which have higher S/N.
 These findings suggest that this DL method can be efficiently applied to novel broad band surveys, such as LSST, with no much gain to the use of more bands in the current range of magnitudes and redshifts.
 
 Another possibility would be to incorporate the current models in a Neural ordinary differential equations (NODE) approach as in \citet{Gupta2020}. In fact, NODE is an efficient way to train a DNN that does not require large data set for training and we may find a gain using more bands, when implementing them in the pre-training.

 \subsection{Deep Learning performance and the relevance of traditional machine learning algorithms}
 
 Deep Learning exceeds human performance in several computer vision problems, including classification of astronomical sources \cite[see, e.g.][]{challenge01,challenge02,ILSVRC15}. In particular, the validation set, not used for training the network weights, reached an accuracy of $\sim 98 \pm 1\%$, which leaves little room for improvement or any measurable bias towards magnitude or redshifts in the current sample. In fact, these results are comparable to other works using a similar DNN architecture. For instance,  \citet{Kalvankar2020} found $92.58\%$ in accuracy for ETG and Sp classification, while \citet{Cheng2020} found $99\%$ in $2,800$ DES sample and \citet{Farias2020} found $\sim 98 \% $ in a SDSS sample of jpg images. In \citet{Kalvankar2020} an extensive multiclass classification is explored. In such a regime, we found more ambiguity between classes and fewer examples per class. For this fine grained classification, it might be the case where more bands could be more advantageous. With fewer examples per class (sometimes orders of magnitude),  this is also a regime where ML with morphological features  other than DL, as the ones from MFMTK, could find competitive results. In fact, Figure \ref{fig:evo_C_H} present how the morphological features evolve in different bands in the two groups classified by the method presented in this paper. It is noteworthy the median separation between ETG and Sp. The figure presents visual prospect on how the $12$ band morphological features could also be exploited for the current binary classification using an algorithm with lower computational complexity, i.e. easily scalable for Big Data regime. We leave this study, including multiclass approach, for a future contribution (Lucatelli et al, in prep.)

 \subsection{Agreement between Deep Learning classification intuition given in morphological features}
 
The use of DL for imaging recognition extends to way more than astrophysics. Consequently, this method has rapidly advanced and improved in the last years, providing a database for pre-training, ideas and tools that have been largely applied in astronomy, resulting in a wealth of publications (see for example \citet{Dominguez2018, Barchi2020,tuccillo2016deep,khalifa2018deep,dieleman2015rotation,zhu2019galaxy,dai2018galaxy}). \citet{Barchi2020} applied Ml and DL techniques to the Sloan Digital Sky Survey Data Release 7 (SDSS-DR7), providing a catalogue of  670,560 galaxies. They achieve a 99\% accuracy on average when classifying galaxies into two classes (ETG's and Sp's). This result compares well with what obtained in this work, somehow not surprisingly (or maybe reassuring) since they make use of the same training and validation set (i.e. GZ1) and of a similar method based on DL. To deeply explore the results of these two methods, we can compare Figure \ref{fig:compDeep}  of this work with figures 11 and 12 in \citet{Barchi2020}. These figures show which T-Type, as obtained by \citet{Dominguez2018} and \citet{Nair2010}, corresponds to ETG and Sp galaxies respectively. In theory, ETG and S0 --  early-type galaxies -- should have a T-Type $\leq 0$, while Sp -- late-type galaxies -- should present a T-Type > 0. On average, both works recover such behaviour, especially when compared with the visual classification of \citet{Nair2010} (note that in this work the number statistics in this comparison are lower than in \citealp{Barchi2020}). Interestingly, when comparing with the automated classification of \citet{Dominguez2018}, it is clear that in this work the E galaxies class is better defined (i.e. there are almost no galaxies classified as ETG with a T-Type $> 0$). \citet{Barchi2020} obtains a similar result when selecting only galaxies with a low probability of being S0 galaxies (as defined in \citet{Dominguez2018}). The novel DL algorithm presented in this work is extremely accurate in identifying ETG galaxies. On the other side, some of the galaxies classified as Sp have a T-Type $\le 0$ in \citet{Dominguez2018}, maybe caused by contamination from S0 galaxies. It is hard to compare these results since no code carries the truth, but it is interesting to notice that the main cause of confusion in both the classifications is generally occurring for T-Type $\simeq$ 0 (i.e., where the S0 class lies). In this work, as in previous works \citep{bamford2009galaxy,lintott2008,lintott2011}, S0 galaxies are incorporated to the early-type galaxy class together with the ETG galaxies, since both classes of objects do not present spiral arms. Yet, low resolution or faint images of Sp galaxies can be easily misclassified as S0 galaxies and, therefore, as early-type galaxies.  A morphological classification that explicitly considers this third group of objects is necessary to overcome this problem (Lucatelli et al in prep).

\subsection{Future challenges for Deep Learning morphological classification}

Despite the high performance results, there still open challenges for the DL approach in morphological classification. In particular, DNNs are known to find non trivial solutions which might be hard to interpret \citep{ribeiro2016model,lundberg2017unified}. Thus, it is worth exploring the behaviour of such DNN approach in the survey limits, i.e., in the faint and high redshift end, where humans or measurable morphological features fail, to see if and how the DNN could contribute beyond what one would expect. Other major challenge is to explore the range of small Petrosian radius.

This method will subsequently be applied to the whole S-PLUS survey, providing a morphological classification of galaxies for 8000 $\rm deg^2$ of the southern hemisphere. We are also currently investigating the use of this method to obtain detailed features such as the ones defined in GZ2 such as bulge, spiral arms and bars among others. These features are  more challenging not only because they represent a bigger variety, but also because these are non mutually exclusive classes. This case presents an interesting application of the S-PLUS filters as some features can be better constrained in different filters. Combining galaxy morphology with other data products of the S-PLUS survey, such as stellar population properties, as obtained through SED fitting of the 5 broad and 7 narrow bands, as well as environment measures, recovered via the precise photo-z determinations ($\delta z \simeq 0.03$), will allow to map the large scale structure of the local Universe and probe the dependence on mass and environment of galaxy evolution.  
Our results enhance the use of pre-trained weights. Presently, due to ImageNet, we limit the pre-training to the 3-band case. However, as we produced models in multiband datasets, one could use those results in the advantage of  pre-trained models for transfer learning using more than three bands in other surveys such LSST \citep{LSST20}, Euclid \citep{EUCLID}, Nancy Grace Roman Space Telescope \citep{WFIRST} among others. In particular, the 8 narrow band model could be used as a starting point for narrow band surveys such as J-PAS \citep{JPAS} and J-PLUS \citep{JPLUS}. By using the current pre-trained models in those surveys it would be possible to evaluate if pre-training in more bands can achieve a gain in a different range of magnitudes and redshifts. Thus, we release out DL models along with our catalogue as a value-added product for the astronomical community.

\section*{Data Availability}
We make the morphological catalogues developed and some extra comparison panels publicly available in \url{https://github.com/cdebom/splus_morph}. The trained Deep Learning models are also public and can be downloaded in \url{https://doi.org/10.5281/zenodo.4891060}.

\section*{Acknowledgements}

The S-PLUS project, including the T80-South robotic telescope and the S-PLUS scientific survey, was founded as a partnership between the Funda\c{c}\~{a}o de Amparo \`{a} Pesquisa do Estado de S\~{a}o Paulo (FAPESP), the Observat\'{o}rio Nacional (ON), the Federal University of Sergipe (UFS), and the Federal University of Santa Catarina (UFSC), with important financial and practical contributions from other collaborating institutes in Brazil, Chile (Universidad de La Serena), and Spain (Centro de Estudios de F\'{\i}sica del Cosmos de Arag\'{o}n, CEFCA). We further acknowledge financial support from the São Paulo Research Foundation (FAPESP), the Brazilian National Research Council (CNPq), the Coordination for the Improvement of Higher Education Personnel (CAPES), the Carlos Chagas Filho Rio de Janeiro State Research Foundation (FAPERJ) and the Brazilian Innovation Agency (FINEP).

The authors who are members of the S-PLUS collaboration are grateful for the contributions from CTIO staff in helping in the construction, commissioning and maintenance of the T80-South telescope and camera. We are also indebted to Rene Laporte and INPE, as well as Keith Taylor, for their important contributions to the project. From CEFCA, we particularly would like to thank Antonio Mar\'{i}n-Franch for his invaluable contributions in the early phases of the project, David Crist{\'o}bal-Hornillos and his team for their help with the installation of the data reduction package \textsc{jype} version 0.9.9, C\'{e}sar \'{I}\~{n}iguez for providing 2D measurements of the filter transmissions, and all other staff members for their support with various aspects of the project.

CMdO and LSJ acknowledge funding for this work from FAPESP grants 2019/26492-3, 2019/11910-4, 2019/10923-5 and 2009/54202-8. GS, CMdO and LS acknowledge  support, respectively, from CNPq grants 309209/2019-6, 115795/2020-0 and 304819/201794. NM acknowledges the University of São Paulo PUB grant 83-1 of 2020.  
A. C. acknowledge the financial support provided by CAPES.

The authors made use of multi GPU Sci-Mind machines developed and tested for Artificial Intelligence and would like to thank P. Russano and P. Souza Pereira for all the support in infrastructure matters. The authors would like to thank R. C. T. de Souza, C.E. Barbosa, A. L. Chies-Santos, H.Farias, D.Ortiz, M.Jaque Arancibia for useful suggestions and comments.

The authors made use and acknowledge TOPCAT\footnote{\url{http://www.starlink.ac.uk/topcat/ (TOPCAT)}} \citep{2005ASPC..347...29T}   tool to analyse the data.

 For the panel in the appendix the authors made use of small cut outs images from the Legacy Survey. The Legacy Surveys consist of three individual and complementary projects: the Dark Energy Camera Legacy Survey (DECaLS; Proposal ID \#2014B-0404; PIs: David Schlegel and Arjun Dey), the Beijing-Arizona Sky Survey (BASS; NOAO Prop. ID \#2015A-0801; PIs: Zhou Xu and Xiaohui Fan), and the Mayall z-band Legacy Survey (MzLS; Prop. ID \#2016A-0453; PI: Arjun Dey). DECaLS, BASS and MzLS together include data obtained, respectively, at the Blanco telescope, Cerro Tololo Inter-American Observatory, NSF’s NOIRLab; the Bok telescope, Steward Observatory, University of Arizona; and the Mayall telescope, Kitt Peak National Observatory, NOIRLab. The Legacy Surveys project is honored to be permitted to conduct astronomical research on Iolkam Du’ag (Kitt Peak), a mountain with particular significance to the Tohono O’odham Nation.
 
\clearpage

\newpage
\clearpage

\newpage
\clearpage

\bibliographystyle{mnras} 
\bibliography{bibliografia}

\appendix

\section{Notes on the Ambiguous sample}
\label{appendix}
\begin{figure}
\centering
\includegraphics[width=0.9\linewidth]{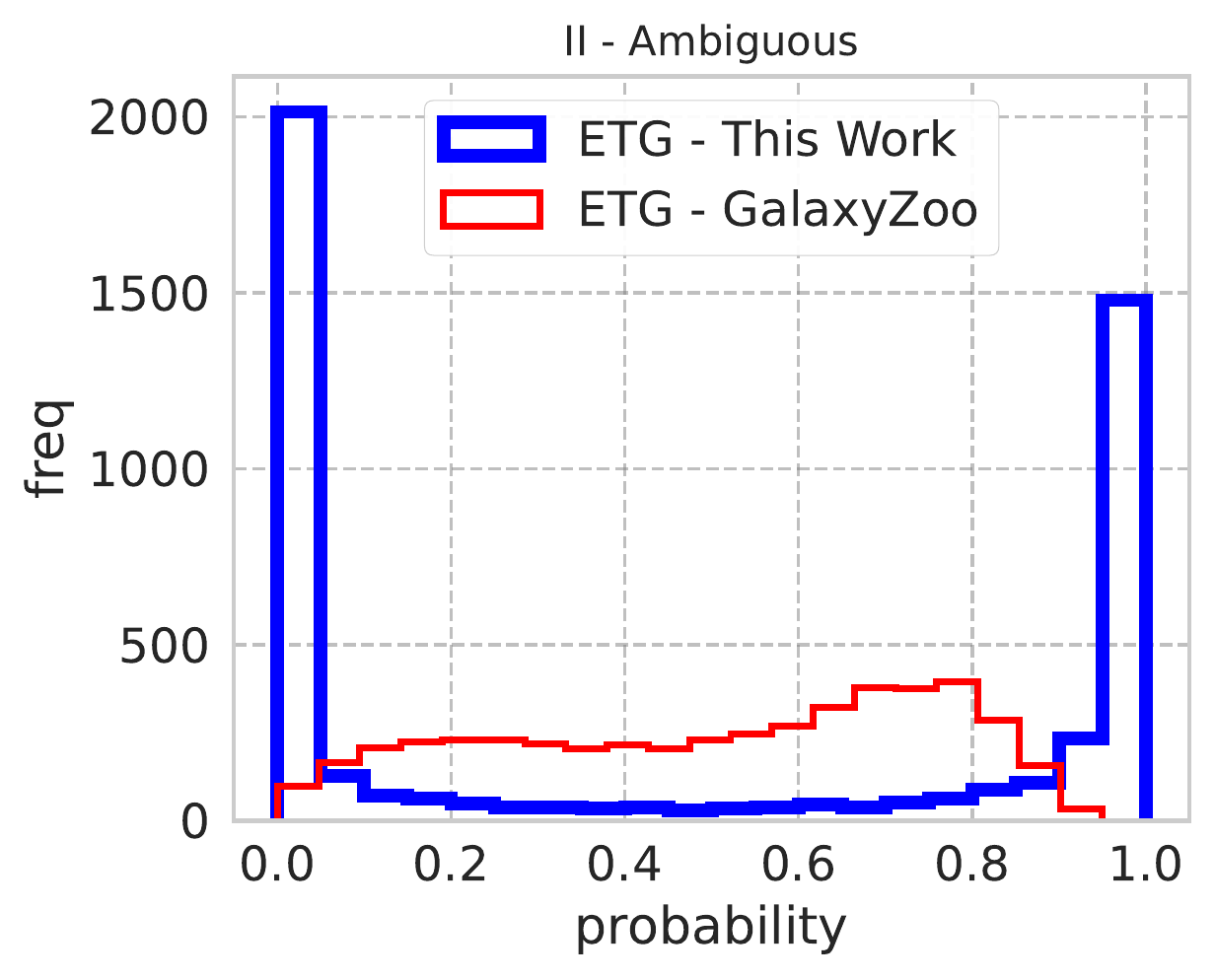}\\
\includegraphics[width=0.9\linewidth]{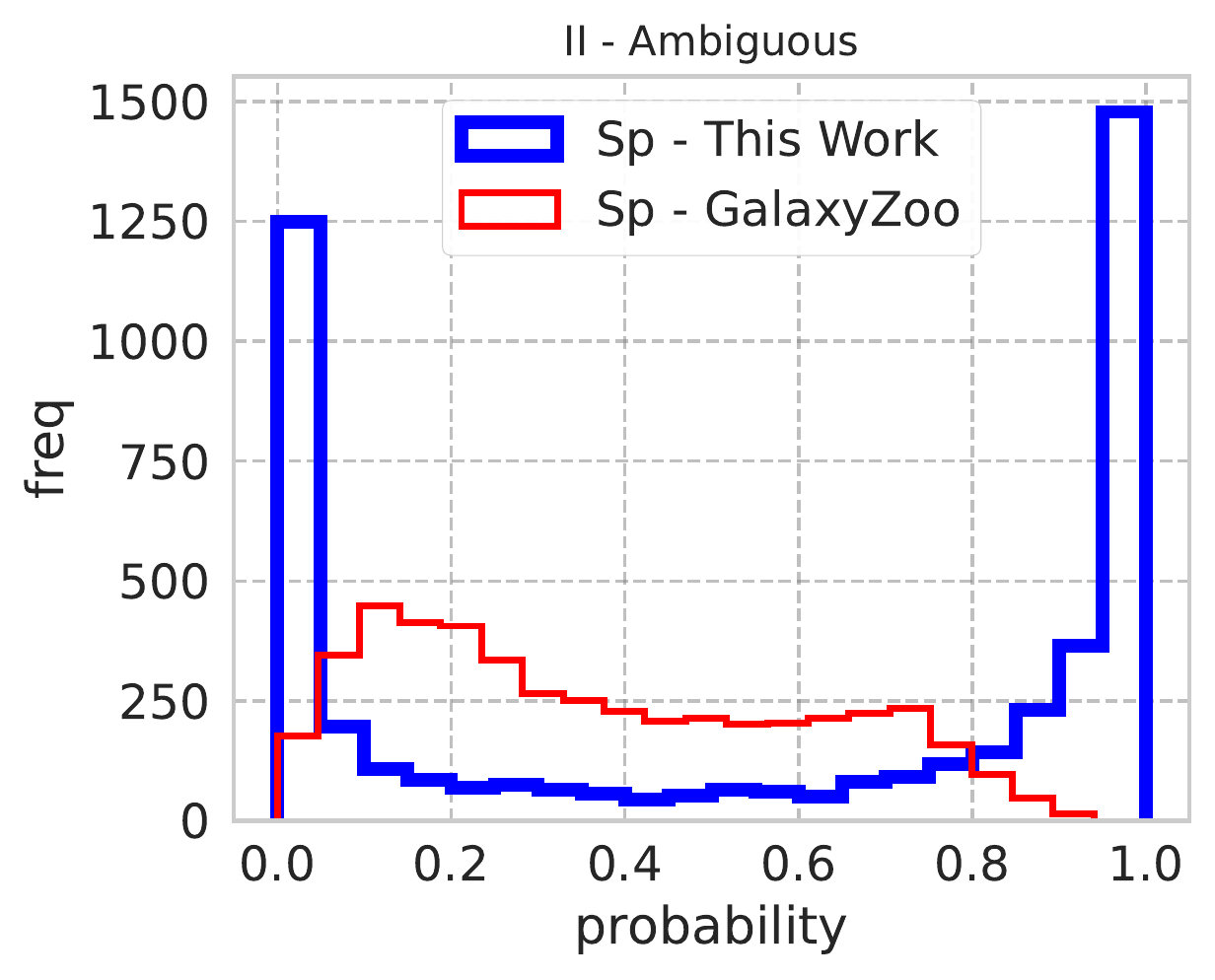}
    \caption{
    \label{fig:amb_comp_prob}
    Distribution of probabilities (scores) for galaxies to be ETG or Sp provided by the Neural Network (blue) in comparison with the debiased likelihood from GZ1 (red).
    }
\end{figure}

The debiased classification published in Table 2 of the GZ1 data release provides a debiased likelihood of a galaxy being a spiral or an elliptical and a morphology flag \citet{lintott2011}, using a binary classification (0=false,1=true) that represents a final catalogue of ETG and Sp likelihood\footnote{For further information we also refer to the data release notes in \url{https://data.galaxyzoo.org/}}. The deep learning technique described in this contribution used the morphology flag instead of the debiased likelihood from Galaxy Zoo. We discussed this choice in Section \ref{subsec:GZ1_training}. While all galaxies in Table 2 of GZ1 data release have a debiased likelihood of being elliptical or spiral galaxies, they might have a flag zero in both elliptical and spiral morphology, see \citet{bamford2009galaxy,lintott2011} for more details.
Nevertheless, we can compare the probability of being a spiral or a lenticular galaxy obtained in this work using a DL method with the debiased likelihood provided in \citet{lintott2011}, as shown in Figure \ref{fig:amb_comp_prob}.

\begin{figure}
\centering
\includegraphics[width=0.95\linewidth]{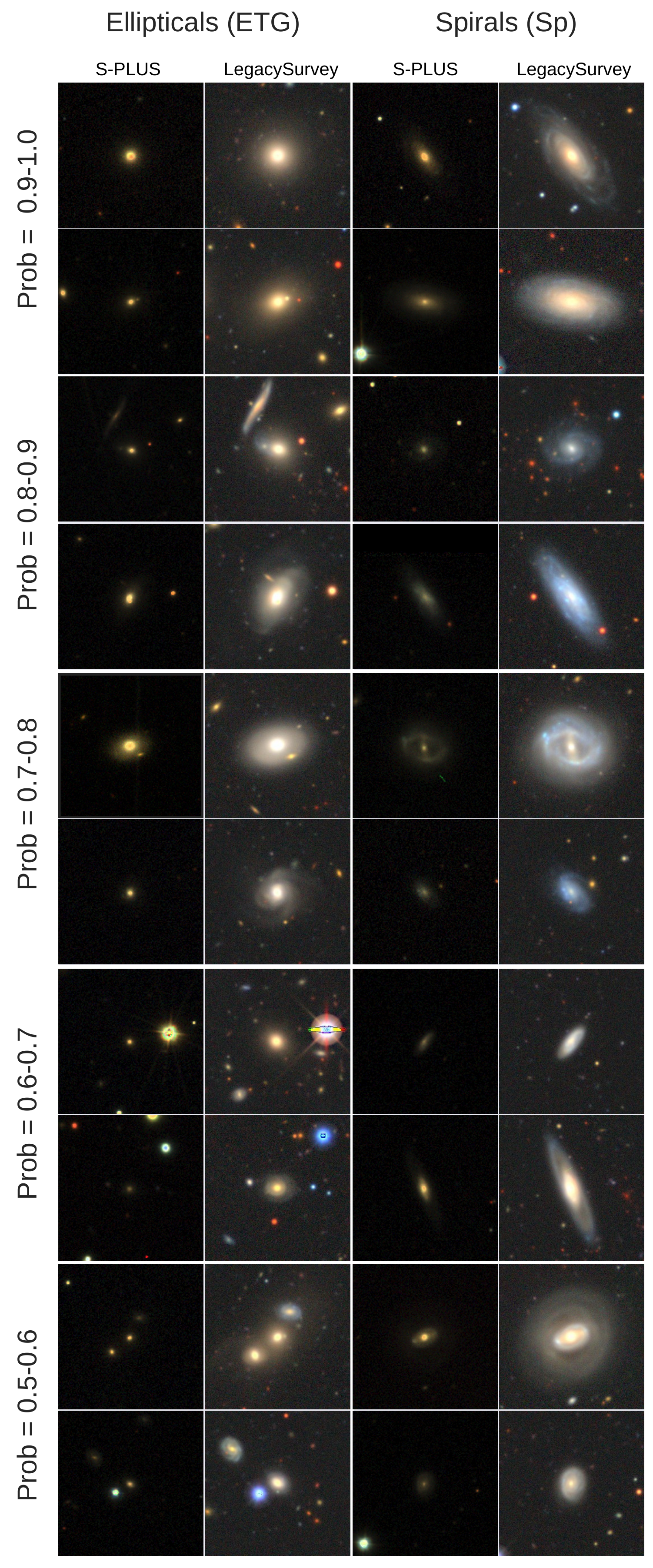}\\
    \caption{
    \label{fig:amb_comp}
    Example of color images showing  ETG galaxies {\it (left panels)} and a Sp galaxy {\it(right panels)} for S-PLUS and Legacy Survey. \\
    }
\end{figure}

The debiased likelihood of the ambiguous set presents a smooth distribution. It is worth noticing that the threshold cut of $0.8$ presented in \citep{bamford2009galaxy,lintott2011} is not equivalent to the flag type used in this work, which includes objects with probability higher than $0.8$ of being either Sp or E according to GZ1, resulting in a morphology flag equal to 0 in both classes. 
On the other side, the result of the DL method (the blue histogram in Figure \ref{fig:amb_comp}) shows a clear bimodality and a net classification between elliptical and spiral galaxies. To check if the Deep Learning method is overconfident, we inspected by eye all the galaxies belonging to the sub-sample II, comparing them with the deepest Legacy Survey \citep{dey2019overview,zou2017project} data. We present an example of such a comparison in Figure \ref{fig:amb_comp}.

It is interesting to see that, in general, galaxies classified as spiral by the DL method present clear spiral arms in the Legacy data, even if these arms are barely visible in S-PLUS images. Note that the depth of the S-PLUS survey is slightly higher than the standard SDSS depth, which characterises GZ1 galaxies. 
On the other side, some galaxies classified as ETGs by the DL method present hints of spiral arms in the Legacy data. Clearly, the S-PLUS imaging only detects the central part of the galaxy light, where the bulge is dominant, revealing the limitations due to the data.

\end{document}